\documentclass[12pt]{article}

\usepackage{fullpage}
\usepackage{amsmath,amssymb,amsthm,amscd,graphicx}
\usepackage{hyperref}
\usepackage{bbold}
\usepackage{amsfonts}
\usepackage{stmaryrd}
\SetSymbolFont{stmry}{bold}{U}{stmry}{m}{n}
\usepackage{caption}
\usepackage{subcaption}
\usepackage{epigraph}
\usepackage[indent]{parskip}
\usepackage{geometry}
\usepackage{tikz}
\usepackage{bm}

\numberwithin{equation}{section}

\renewcommand\hat\widehat
\renewcommand\tilde\widetilde
\newcommand{\bea}{\begin{eqnarray}}
\newcommand{\eea}{\end{eqnarray}}
\newcommand{\be}{\begin{equation}}
\newcommand{\ee}{\end{equation}}
\newcommand{\nn}{\nonumber}

\def\+{{+\!\!\!+}}

\theoremstyle{definition}

\begin{document}
\begin{flushright}Imperial-TP-2024-CH-03
\\
UUITP-05/24
\end{flushright}
\
 \\[12mm] 

\begin{center} \Large
{\bf  $N=(2,2)$ superfields and geometry revisited}
 \\[15mm] \normalsize
{\bf Chris Hull${}^{a}$  and Maxim Zabzine${}^{b,c}$}\\
[8mm]
{\small\it ${}^a$ The Blackett Laboratory, Imperial College London, Prince Consort Road, \\
London SW7 2AZ, United Kingdom}\\
{\small\it ${}^b$Department of Physics and Astronomy, Uppsala University,\\ Box 516, SE-75120 Uppsala, Sweden}\\
{\small\it ${}^c$ Centre for Geometry and Physics, Uppsala University,\\ Box 480, SE-75106 Uppsala, Sweden }\\

\end{center}
\vspace{10mm}
\begin{abstract}
 We take a fresh look at the relation between generalised K\"ahler geometry and $N=(2,2)$ supersymmetric sigma models in  two dimensions formulated in terms of $(2,2)$ superfields. 
 Dual formulations in terms of different kinds of superfield are combined to give  a formulation with a  doubled target space and both the original superfield and the dual  superfield. For K\"ahler geometry, we show that this doubled geometry is Donaldson's deformation of the holomorphic cotangent bundle of the original K\"ahler manifold. This doubled formulation gives an elegant geometric reformulation of the equations of motion.  We   interpret the equations of motion
   as the intersection of two Lagrangian submanifolds
   (or of a Lagrangian submanifold with an isotropic one)
    in the infinite dimensional  symplectic supermanifold 
    which is the analogue  of phase space. 
    We then consider further extensions of this formalism, including one in which the geometry is quadrupled, and discuss their
geometry. 
\end{abstract}

\eject

\tableofcontents

\eject

\section{Introduction}

\epigraph{Superfields are smarter than we are!}{Martin Ro\v{c}ek told us this many times}

Two dimensional $N=(2,2)$ supersymmetric non-linear sigma models play a prominent role in string theory and in 
 mathematical physics,
 with a beautiful relation between   supersymmetry and the geometry of the target space.
   Zumino  \cite{Zumino:1979et}  was the first  to  
 realise, in 1979, 
that $N=(2,2)$ supersymmetry constrains the target space of a non-linear sigma-model to be K\"ahler.
Chiral superfields played a prominent role  and the $N=(2,2)$ Lagrangian density was identified with the
   K\"ahler potential.   In \cite{Gates:1984nk}, this was generalised to the case of sigma-models with Wess-Zumino (WZ) term and it was found that the general $N=(2,2)$ models were described in terms of a bihermitian geometry. This  geometry is now called generalised K\"ahler geometry.  In \cite{Gates:1984nk}, a   $N=(2,2)$ superfield formulation was found   for a special class of generalised K\"ahler geometries (those with commuting complex structures that were simultaneously integrable)
   and it involved both chiral and twisted-chiral superfields. 
 A further kind of superfield -- the semichiral superfield -- was introduced in \cite{Buscher:1987uw} and in \cite{Lindstrom:2005zr} the full $N=(2,2)$ superfield 
 description of generalised K\"ahler geometry was found (subject to a certain regularity assumption) and this superfield description was given a
  geometric realisation in terms of local coordinates (which are associated 
    with the various  superfields). For this general model, the $N=(2,2)$ Lagrangian density is given in terms of single 
   function, the generalised K\"ahler potential, which encodes all local geometry.  
   
  In parallel, there were interesting  developments in mathematics,  starting with the work of Hitchin \cite{Hitchin:2003cxu}   who introduced the notion 
 of generalised complex structure.
    Gualtieri, in his PhD thesis \cite{Gualtieri:2003dx}, introduced the notion of generalised K\"ahler geometry and showed that this is the same as the  bihermitian geometry of  \cite{Gates:1984nk} (see also
     \cite{Gualtieri:2014kja}).
Later, a   global interpretation of the generalised K\"ahler potential was given  in \cite{Bischoff:2018kzk} for the case of a generalised K\"ahler structure of 
 symplectic type.

 In this paper, we return to the superfield description of the  $N=(2,2)$ sigma model and find new geometrical structure.
 Strikingly, it is the superfield structure that suggests the new geometry.

 We start with the K\" ahler geometry of the $N=(2,2)$ sigma-model without WZ term.
 As well as the usual formulation in terms of chiral superfields, there is a dual formulation in terms of complex linear superfields.\footnote{As we will discuss later,  there is, strictly speaking, only a full global dual formulation for a special class of target spaces.} 
 In one formulation, the coordinates appear as the lowest component of chiral superfields, in the other as the lowest component of complex linear superfields.
 We reformulate the general K\"ahler sigma-model in a \lq doubled target space' with both chiral superfields and  complex linear superfields. We then impose a relation between the two types of superfield such that the superspace constraint on  either one implies the field equation of the other. This gives an elegant set of equations that captures the classical structure of the model.
 Moroever, for those target spaces for which there are two dual formulations, this 
    makes the duality  manifest.
We show that the geometry of the doubled space is that of Donaldson's deformation of the holomorphic cotangent bundle of the original K\"ahler manifold, while the relation between the two types of superfield constructs the original space as a Lagrangian submanifold of the deformed   cotangent bundle.

Next, we lift this structure  to the space of unconstrained maps from $(2,2)$ superspace to the deformed   cotangent bundle, which 
might be thought of as a kind of  phase space.
We show that this has a holomorphic symplectic structure. We show that the superspace constraints correspond to restricting to a holomorphic isotropic submanifold,  so that  the 
  sigma model is specified by the intersection of an isotropic submanifold and a 
 Lagrangian  submanifold. We also give an extended formulation in which thesubmanifold specified by the superfield constraints is also Lagrangian.

We then extend this formulation to the case of $N=(2,2)$ sigma-models with Wess-Zumino  term and their generalised K\"ahler geometry.
We find an enlarged formulation in which, as well as  the chiral, twisted-chiral and semichiral superfields, we include the dual superfields.
The model is specified by a constraint that selects a submanifold of this enlarged space and implies the field equations.
We discuss the symplectic geometry of this and show that the model can be formulated as the intersection of two submanifolds 
of the space of supermaps to this enlarged phase space.
(These submanifolds are either both Lagrangian, or one is Lagrangian and one is isotropic, depending on the choice of extended space.)
We discuss several different enlarged spaces, each of which gives an extended formulation of the original sigma models.

  In our discussion, we combine the superfield description from \cite{Lindstrom:2005zr} with ideas coming from dualities \cite{Gates:1983nr} and global considerations from 
  \cite{Hull:2008vw}.  The superfield formalism imposes a very rigid structure on field transformations both at local and global levels and as a result constrains the kind of transition functions that are permitted.
  The gerbe structure found in \cite{Hull:2008vw} underpins the geometry, and we find that in some cases it leads to interesting spaces that      might be considered as  generalisations of  fibre bundles.
 It will be interesting to compare all this  with the global discussion in  \cite{Bischoff:2018kzk} and     the  recent results in \cite{Marco-paper} and  in \cite{jiang}.   We think that the ideas we set out here
              can be further developed and generalised beyond the present paper.      
           
 The paper is organised as follows.  In section \ref{s:kahler} we  review    K\"ahler geometry and the  formulation of the corresponding sigma model in terms of chiral superfields and the dual formulation in terms of complex linear superfields. We then present a doubled formulation in terms of both kinds of superfields and show that the doubled target space is Donaldson's
    deformation of the  holomorphic contangent bundle. In section \ref{s:general} we find a similar doubled formulation of the
   general $N=(2,2)$ sigma model with 
WZ term;     in 
     this section, we present only a local analysis.     
  In section  \ref{s:examples} we discuss the global issues for this construction. 
       Sections \ref{s:extended} and \ref{s:double}  extend the phase space   further in    constructions 
        dictated by the superfield formalism.  In Section  \ref{s:extended} we consider a construction in which  we quadruple some fields 
       and double others,  while in Section \ref{s:double} we offer a democratic formulation in which   all fields are quadrupled. 
             Finally, in section 
  \ref{s:summary} we summarise our results and outline open problems. In  two appendices we collect some basic properties of the  $N=(2,2)$ superfield 
   formalism and review some features of   generalised K\"ahler geometry  and its gerbe structure.

\section{K\"ahler Geometry and the $N=2$ Supersymmetric Sigma Model}\label{s:kahler}

In this section we consider  the well known case of K\"ahler geometry and the corresponding $N=(2,2)$ supersymmetric sigma models.
 We first review the local structure of the sigma model and its dual formulation.
 We then provide a  system of equations combining  the equations of motion and the superspace constraints in a unified way. This system has a manifest duality that interchanges field equations with constraints.  The equations govern superfields taking values in a \lq doubled space', with the original space arising as a Lagrangian submanifold. We show that this doubled space
 is Donaldson's deformed cotangent bundle and this
 facilitates the discussion of  global issues.  Finally, we propose a new geometric  interpretation of the system in an  infinite-dimensional setting.

\subsection{The K\"ahler Sigma Model and its Dual Formulation}

The  $N=(2,2)$ sigma model with K\"ahler target $M$ is a theory of maps
\be
   {\cal N} = \{ \phi :\mathbb{R}^{2|4}\longrightarrow {M} \}
  \ee
  from $N=(2,2)$ superspace $\mathbb{R}^{2|4}$ with coordinates $(x^m,\theta^\alpha  )$ (where $\alpha=+,-$ are spinor indices and $\theta^\alpha  $ are complex) to the manifold $M$ with local complex coordinates $(\phi^{a},\bar{\phi}^{\bar{a}})$. The maps are locally specified by coordinate maps $\phi^{a}(x^m,\theta^\alpha )$ which are constrained to be chiral
\be
 \bar{ \mathbb{D}}_{\pm} \phi^{a} =0~,~~~~~~
 \mathbb{D}_{\pm} \bar{\phi}^{\bar{a}} =0
 \ee
and the action is
\be\label{action-Kah}
 S = \int d^2x~ d^4 \theta ~K (\phi^{a}, \bar{\phi}^{\bar{a}})~,
\ee
 where  
 the real function $K (\phi^{a}, \bar{\phi}^{\bar{a}})$
 is the K\"ahler potential. See Appendix A for details of our superspace conventions and notation.
 
 The sigma-model has a remarkable dual formulation in terms of complex linear superfields,  given in  
the superspace textbook \cite{Gates:1983nr} for the four-dimensional sigma model (see the discussion around eqs (4.5.10a,b)).
Complex linear superfields were introduced in \cite{GATES1981389} and further discussion of this dual formulation can be found in \cite{DEO1985187} for the four-dimensional sigma model
 and in  \cite{Penati:1998yt} for the two-dimensional sigma model.
First, note that the action (\ref{action-Kah}) can be rewritten
in a  first-order form as
 \be\label{dual-Kah}
  S_{\rm dual} = \int d^2x~ d^4 \theta ~\Big [ K (\Phi^b, \bar{\Phi}^{\bar{b}}) - \Sigma_a \Phi^a - \bar{\Sigma}_{\bar{a}} \bar{\Phi}^{\bar{a}} \Big ] ~,
 \ee
 where $\Phi^b$ are unconstrained complex superfields and $\Sigma_a$  are complex linear superfields satisfying  the constraint
 \be
 \bar{\mathbb{D}}_+  \bar{\mathbb{D}}_- \Sigma_a=0
 \ee
  ($\bar{\Sigma}_{\bar{a}}$ are the  complex conjugate superfields). 
 The field equation for $\Sigma_a$
 imposes the chirality constraint $ \bar{ \mathbb{D}}_{\pm} \Phi^{a} =0$ so that 
 integrating out the $\Sigma_a$'s recovers the original action (\ref{action-Kah}) 
 (on identifying the constrained $\Phi^{a} $ with the chiral superfields $\phi^{a} $).
 On the other hand, the field equation for $\Phi^a$ is
  \be
  \label{sigiss}
  \Sigma_a = \frac{\partial K}{\partial \Phi^a}
  \ee
  giving $  \Sigma_a $ as a function of $\Phi^a, \bar{\Phi}^{\bar{a}}$. If this can be inverted to give $\Phi^a$ as a function of $\Sigma_a,\bar{\Sigma}_{\bar{a}}$, then integrating  
 out $\Phi^a, \bar{\Phi}^{\bar{a}}$ sets $\Phi^a$  to be the function
  $\Phi^a(\Sigma_a, \bar{\Sigma}_{\bar{a}})$
  giving the
   dual formulation 
    \be\label{dual-second}
   S = \int d^2x~ d^4 \theta ~\tilde{K} (\Sigma_a, \bar{\Sigma}_{\bar{a}})~,
   \ee
    where $\tilde{K}$ is the Legendre transform of $K$:
    \be
    \tilde{K}=K (\Phi^b, \bar{\Phi}^{\bar{b}}) - \Sigma_a \Phi^a - \bar{\Sigma}_{\bar{a}} \bar{\Phi}^{\bar{a}}~,
    \ee
    and (\ref{sigiss}) is inverted and used to write the RHS as a function of $\Sigma_a, \bar{\Sigma}_{\bar{a}}$.
    For
    this to be a good duality 
    requires
    that the function $  \Sigma_a (\Phi^a, \bar{\Phi}^{\bar{a}})$ is invertible and that the Legendre function is well-defined (not multivalued) and so in particular needs $K$ to be a convex function. The general case in which it is not convex is interesting and will be discussed elsewhere.

Now the dual action (\ref{dual-second}) can itself be written in a first-order form as
     \be\label{another-dual-Kah}
      \tilde{S}_{\rm dual} = \int d^2x~ d^4 \theta ~\Big [ \tilde{K} (\Phi_b, \bar{\Phi}_{\bar{b}}) - \phi^a \Phi_a - \bar{\phi}^{\bar{a}} \bar{\Phi}_{\bar{a}} \Big ] ~,
    \ee
     where the  $\Phi_b$ are complex unconstrained superfields and the $\phi^a$ are chiral superfields. 
 The $\phi$ equation of motion imposes the constraint that  $\Phi_a$  is a complex linear superfield,  $ \bar{\mathbb{D}}_+  \bar{\mathbb{D}}_- \Phi_a=0$, 
 so that integrating out $\phi$ recovers 
  the action (\ref{dual-second}) (on identifying the constrained $\Phi_{a} $ with the complex linear superfield $\Sigma_a $). On the other hand, the $\Phi$ field equation gives
  \be
  \phi^a = \frac{\partial \tilde{K}}{\partial \Sigma_a}~.
  \ee
  Integrating out the fields $\Phi_a$ requires solving this to determine $\Sigma$ as a function of
  $\phi^{a}, \bar{\phi}^{\bar{a}}$
  and substituting in the action (\ref{another-dual-Kah}) to arrive
  back  at the original action 
       (\ref{action-Kah}) (for this to be well-defined, $\tilde K$ should be convex).

 \subsection{ Lagrangian System for the K\"ahler Sigma Model}

The equations of motion following from the action (\ref{dual-Kah}) together with the superfield constraints can be recast as  the following system of equations
 \be\label{Kah-eqsmot}
\boxed{ \Sigma_a = \frac{\partial K}{\partial \phi^a}~,~~~~ 
 \bar{\mathbb{D}}_+  \bar{\mathbb{D}}_- \Sigma_a=0~,~~~~
  \bar{ \mathbb{D}}_{\pm} \phi^a =0~,}
 \ee
 where $K(\phi ,\bar \phi)$ is the K\"ahler potential.
 Indeed, the first two equations combine to give
 \be
 { \bar{\mathbb{D}}}^2  \frac{\partial K}{\partial \phi^a}=0~,
 \ee
 which are precisely the equations of motion following from (\ref{action-Kah}).
      Alternatively,   the equations of motion  derived from the dual action (\ref{another-dual-Kah})   can be recast as  the following system 
   \be\label{Kah-eqsmot-dual}
\boxed{ \phi^a = \frac{\partial \tilde{K}}{\partial \Sigma_a}~,~~~~ 
 \bar{\mathbb{D}}_+  \bar{\mathbb{D}}_- \Sigma_a=0~,~~~~
  \bar{ \mathbb{D}}_{\pm} \phi^a =0~,}
 \ee
 where $\tilde{K}(\Sigma,\bar \Sigma)$ is the dual potential. Then the first and last equation combine to give the field equation for $\Sigma$.

 For both systems (\ref{Kah-eqsmot}) and (\ref{Kah-eqsmot-dual}) we have  maps from superspace to a space with complex coordinates  $(\phi^a,\Sigma_a)$ satisfying the constraints $ \bar{\mathbb{D}}_+  \bar{\mathbb{D}}_- \Sigma_a=0$, $
  \bar{ \mathbb{D}}_{\pm} \phi^a =0$ and the model is defined by restricting to a subspace 
  defined by the $d$ complex equations (for a manifold of complex dimension $d$) $\Sigma_a = \frac{\partial K}{\partial \phi^a}$ or $\phi^a = \frac{\partial \tilde{K}}{\partial \Sigma_a}$.

 The two systems of equations (\ref{Kah-eqsmot}) and (\ref{Kah-eqsmot-dual})  are equivalent provided that the Legendre transform  
  between $K$ and $\tilde{K}$ is well-defined. 
 This duality between the two systems  interchanges constraints with field equations.
In the first system (\ref{Kah-eqsmot}), we can regard $\phi$  as the fundamental field with $
  \bar{ \mathbb{D}}_{\pm} \phi^a =0$ viewed as a constraint and the other two equations combining to give the field equation for $\phi$ that follows from the action (\ref{action-Kah}). For the dual system (\ref{Kah-eqsmot-dual}) 
  we can regard $\Sigma$  as the fundamental field with $ \bar{\mathbb{D}}_+  \bar{\mathbb{D}}_- \Sigma_a=0$ viewed as a constraint and the other two equations combining to give the field equation for $\Sigma$ that follows from the action (\ref{dual-second}).

 For  example, in the simple   case of a flat target space, the  K\"ahler potential is quadratic, $K = \frac{1}{2} \sum_a \phi^a \bar{\phi}^{\bar{a}}$. Then $\Sigma_a = \frac{\partial K}{\partial \phi^a}$ gives $ \Sigma_a= \bar{\phi}^{\bar{a}}$
 so that the complex linear constraint on $\Sigma$, which is $ \bar{\mathbb{D}}_+  \bar{\mathbb{D}}_- \Sigma_a=0$, gives the field equation for a free chiral superfield,
 $ \bar{\mathbb{D}}_+  \bar{\mathbb{D}}_- \bar{\phi}^{\bar{a}}=0$.
 The dual potential is $\tilde K = \frac{1}{2} \sum_a \Sigma_a \bar{\Sigma}_{\bar{a}}$
  so that we again obtain
  $ \Sigma_a= \bar{\phi}^{\bar{a}}$ and now the chiral constraint on $\phi$ implies the complex linear constraint on $\Sigma$, i.e.\ $ \bar{\mathbb{D}}_+  \bar{\mathbb{D}}_- \Sigma_a=0$.

    This duality becomes rather   subtle   if $K$ is not convex as then the Legendre transform  $\tilde{K}$ becomes multivalued and may even be singular.     
     We will not discuss this duality further here and will return to this subject elsewhere.  
     Instead, we will  focus here on the system (\ref{Kah-eqsmot}) which is defined (locally) for any K\"ahler manifold and not assume the existence of a dual potential $\tilde K$.
We now turn to the global formulation of this system for any     K\"ahler manifold.

\subsection{Global Structure}\label{ss:global-kah}

We now look at the global structure of the system (\ref{Kah-eqsmot}).  First we discuss the symmetries of the equations (\ref{Kah-eqsmot}). 
 The chirality conditions allow the field redefinitions $\phi^a\rightarrow {\phi'}^a(\phi)$   corresponding to a holomorphic change 
    of complex coordinates on $M$.  
    Then $\Sigma_a = \frac{\partial K}{\partial \phi^a}$ requires that the
      linear complex superfield   transforms as
    \be\label{diff-sigma}
     \Sigma_a \rightarrow   \Sigma'_a=\frac{\partial \phi^b}{\partial{\phi'}^a}~  \Sigma_b ~.
    \ee
    Note that this  
     is compatible with the superfield structure (see Appendix): if  $ \Sigma_a$
 satisfies the complex linear constraint $ \bar{\mathbb{D}}^2 \Sigma_a=0$, then so does $ \Sigma'_a$.
 Then $ \Sigma_a$ transforms like a holomorphic 1-form and  $(\phi, \Sigma)$ transform under changes of coordinates in the same way as the holomorphic  coordinates
 of the 
cotangent bundle.
However, although       it is tempting to suppose that $(\phi, \Sigma)$ correspond to complex coordinates on the cotangent bundle $T^*M$,
     there is a subtlety related to the way the coordinates $\Sigma$ are glued.
     
     The superspace action  is unchanged if the K\"ahler potential is transformed  by
     \be
     \label{Kshift}
     K\to K'=K+f(\phi) +\bar{f}(\bar{\phi})
     \ee
     for any holomorphic function $f(\phi)$. This is sometimes referred to as  a K\"ahler gauge transformation.
     For this to be consistent with $\Sigma_a = \frac{\partial K}{\partial \phi^a}$, it is necessary that $\Sigma_a$ transforms as
     \be
     \Sigma_a\to \Sigma '_a=\Sigma_a + \frac{\partial}{\partial \phi^a}  f (\phi)
     \ee
     so that $\Sigma$ can be viewed as a connection one-form for these transformations.
     These transformations are important for the global structure, as the glueing between patches is through a K\"ahler gauge transformation  composed with a holomorphic diffeomorphism.
     
     For an open cover of the K\"ahler manifold with (contractible) open sets $U_\alpha$ 
     the K\"ahler structure leads to a locally-defined K\"ahler potential
      $K_\alpha$ on each patch $U_\alpha$.
      On the intersection of two patches $U_\alpha \cap U_\beta$ they are glued by a K\"ahler gauge transformation 
 \be\label{KP-glue}
  K_\alpha (\phi, \bar{\phi}) - K_\beta (\phi, \bar{\phi}) = f_{\alpha\beta} (\phi) + \bar{f}_{\alpha\beta} (\bar{\phi})~
 \ee
 for some holomorphic function $f_{\alpha\beta} $ on $U_\alpha \cap U_\beta$.
  This leads to a well-defined superspace action as   the right hand side vanishes identically under the superspace integral. 
  To preserve the condition $\Sigma_a = \frac{\partial K}{\partial \phi^a}$
 we postulate the following glueing of $\Sigma$'s on the   intersections
  \be\label{shift-sigma}
 ( \Sigma_a )_\beta
 ~=~ (\Sigma_a )_\alpha+ \frac{\partial}{\partial \phi^a}  f_{\alpha\beta} (\phi)  ~~~~~{\rm on}~U_\alpha \cap U_\beta~,
 \ee
  which is again compatible with superfield constraints.  
  The full glueing conditions combine these with a holomorphic diffeomorphism from the coordinates in $U_\alpha$ to those in $U_\beta$.
  In (\ref{KP-glue}),(\ref{shift-sigma}) we 
  suppress the diffeomorphism part of the transition function; this can be thought of as
   expressing all quantities in the relation in terms of the
   same coordinate system.\footnote{We will similarly the suppress the diffeomorphism part of  glueing conditions throughout this paper.}

  The structure  here is a deformation of the holomorphic cotangent bundle and we follow
    Donaldson's construction of this \cite{MR1959581} (see also \cite{Bischoff:2018kzk}).  
    With local complex 
  coordinates $(z^a_\alpha,p_{a\alpha})$ on $T^* U_\alpha$,  we 
  define a 1-form $p_\alpha = p_{a\alpha} dz_\alpha^a$ on each patch $U_\alpha$
   with the holomorphic affine glueing relation for the fibres  over $U_\alpha \cap U_\beta$
     \be \label{KG-momglform}
   p_\beta=p_\alpha+\partial f_{\alpha\beta}~,
   \ee
   which satisfy the cocycle condition. 
   Expressing all terms in the same coordinate system, this can be written as
 \be\label{KG-momgl}
     ( p _a )_\beta
 ~=~ (p _a )_\alpha + \partial_a f_{\alpha\beta}(z)~.
  \ee

 The total space is constructed  from patches of the form $T^* U_\alpha $ glued together using these transition functions.
   That is,  we can define the holomorphic symplectic affine bundle $({\cal Z}, \omega^{(2,0)})$ as 
  \be
   {\cal Z} = \Big ( \coprod\limits_{\alpha} T^* U_\alpha \Big ) / \sim
  \ee
   with the equivalence relation $p_\alpha\sim p_\beta$ defined by (\ref{KG-momgl}).  
    The bundle ${\cal Z}$ inherits a  canonical holomorphic symplectic structure from $T^*M$
  \be\label{def-hol-Kah}
   \omega^{(2,0)}=  dp_a\wedge  dz^a~,
   \ee
which is well-defined under the identification. In other words,  the  transformations between patches (\ref{KG-momgl}) 
 are symplectomorphisms and so the symplectic structure $\omega^{(2,0)}$  is globally defined on  $ {\cal Z} $.
    The choice of K\"ahler potential is encoded in the choice of a global (non-holomorphic) section ${\cal L} : M \rightarrow {\cal Z}$ that is given by
    \be
   p_a dz^a=\partial K~.
   \ee
    This equation specifies a globally defined  submanifold ${\cal L}$ of ${\cal Z}$ that is Lagrangian with respect to  ${\rm Re}(\omega^{(2,0)})$ and symplectic with respect to 
    ${\rm Im} (\omega^{(2,0)})$.
      For a more detailed  discussion we refer the reader to 
   \cite{MR1959581}  and  \cite{Bischoff:2018kzk}.\footnote{Here and in the rest of the paper we follow conventions which allow us 
    minimise the appearance of factors of $i$ in our formulae.
      In some other works different conventions are used in which the roles of ${\rm Re}(\omega^{(2,0)})$ and ${\rm Im} (\omega^{(2,0)})$ are interchanged. We use the same letter for the section and the Lagrangian submanifold since they are two ways of viewing the same submanifold.}

\subsection{Interpretation}

    The system (\ref{Kah-eqsmot}) is then one of supermaps $\mathbb{R}^{2|4}\longrightarrow {\cal L}$
   with $p(x,\theta,\bar \theta)$ satisfying the complex linear constraints and $z(x,\theta,\bar \theta)$ satisfying the chiral constraints.  
    Let us reformulate this observation in geometrical terms.

  We introduce the infinite dimensional space of  unconstrained
   supermaps from $\mathbb{R}^{2|4}$ to the
     holomorphic symplectic affine bundle 
  \be
   {\cal M} = \{ \Phi :\mathbb{R}^{2|4}\longrightarrow {\cal Z} \}
  \ee
    with local holomorphic Darboux coordinates $\Phi^a  , \Phi_a$ on ${\cal Z}$
  with $\Phi^a$ corresponding to complex coordinates on 
    $M$ (referred to as $z^a$ in the last subsection) while $\Phi_a$ are complex coordinates along the fibre
  (referred to as $p_a$ in the last subsection).   On the space of unconstrained supermaps given locally by  $\Phi^a (x,\theta,\bar \theta) , \Phi_a(x,\theta,\bar \theta)$, we denote the exterior derivative  by $\delta$ so that the basic 1-forms are $\delta\Phi^a  , \delta\Phi_a$ and use $\wedge$ for the wedge product. Then  this infinite dimensional space is equipped with the   holomorphic symplectic structure 
  \be\label{hol-Kahlerfields}
 \Omega =  \int d^2x~ d^4 \theta ~ \delta \Phi_a  \wedge \delta \Phi^a~,
  \ee
  which arises from the holomorphic symplectic structure $\omega^{(2,0)}$ (\ref{def-hol-Kah}) on  ${\cal Z}$.
  
  The main idea here is that the system of equations  
      (\ref{Kah-eqsmot}) can be interpreted as the intersection of two infinite dimensional Lagrangian submanifolds of ${\cal M}$.  
    A  Lagrangian submanifold is isotropic (i.e.\ the  symplectic structure restricted to the submanifold vanishes) and satisfies a maximality condition.  
    In the finite dimensional setting,
    a Lagrangian submanifold 
    of a manifold of dimension $2d$ is an isotropic submanifold with dimension $d$. In the infinite dimensional setting,
         a   Lagrangian subspace 
         of a symplectic vector space 
is an isotropic subspace whose  symplectic compliment
that is also isotropic.
 The extension of this to infinite dimensional manifolds 
        is subtle,  but intuitively the maximality of  a Lagrangian submanifold means that if one tries to enlarge the space then the property of being isotropic is lost.

      The first equation (and its complex conjugate) in  (\ref{Kah-eqsmot}) corresponds to 
       \be
        \Phi_a = \frac{\partial K}{\partial \Phi^a}~,
       \ee
   which defines a   real Lagrangian submanifold of ${\cal M}$ with respect to  ${\rm Re} (\Omega)$ (here $K$ is a  real function of $(\Phi^b, \bar{\Phi}^{\bar{b}})$).
    This infinite dimensional Lagrangian submanifold is induced by a finite dimensional Lagrangian submanifold ${\cal L}$ of ${\cal Z}$. $K$ can be promoted to become the generating function of an
     infinite dimensional space, as we show in next subsection. 
   
     On the other hand,
      the superfield constraints in (\ref{Kah-eqsmot}) specify a holomorphic 
      isotropic
        submanifold with respect to 
      $\Omega$.  To see that this submanifold is isotropic we   evaluate the holomorphic symplectic form on the subspace in which the superfields   are constrained
      to obtain
      \be
      \label{subma}
      \Omega|_{ \bar{\mathbb{D}}_{\pm} \Phi^a=0, ~\bar{\mathbb{D}}_+ \bar{\mathbb{D}}_- \Phi_a=0} 
      =  \int d^2x~ d^4 \theta ~ \delta \Phi_a  \wedge \delta \Phi^a  =0~,
      \ee
     where we have used the representation (\ref{reps-super}) for these constrained superfields. The integral vanishes 
      identically (possibly up to boundary terms, but we assume boundary conditions that eliminate these). 
      For example, (\ref{reps-super}) implies that $\delta \Phi^a$, which is  chiral on the constrained submanifold,
      can be written as $\delta \Phi^a=\bar{\mathbb{D}}^2 W^a$ for some $W^a$, so that on integrating by parts the integrand in (\ref{subma}) becomes 
      $W^a  \wedge \bar{\mathbb{D}}^2 \delta \Phi_a$ which vanishes on the submanifold as $\delta \Phi_a$ satisfies the complex linear constraints there. Thus we see that these superfield constraints correspond to restricting  to
     a  holomorphic isotropic submanifold of ${\cal M}$. In the discussion below we  show that 
        the corresponding submanifold in an extended space is actually holomorphic Lagrangian. 
        
     \subsection{Solving the linear constraints  and extended space}  
            
 The complex linear constraint      
               \be
 \bar{\mathbb{D}}_+  \bar{\mathbb{D}}_- \Sigma_a=0
 \ee
is solved by 
\be
\label{sigpsi}
\Sigma_a= \bar{\mathbb{D}}_\alpha \Psi^\alpha_a
\ee
for some unconstrained 
spinor superfields $\Psi^\alpha_a=(\Psi^+_a,\Psi^-_a)$.
Then the first-order action (\ref{dual-Kah}) can be rewritten as
 \be\label{dual-Kah+}
  S_{\rm dual} = \int d^2x~ d^4 \theta ~\Big [ K (\Phi^b, \bar{\Phi}^{\bar{b}}) + \Psi^\alpha_a \bar{\mathbb{D}}_\alpha \Phi^a 
  +\bar \Psi^\alpha_{\bar a}  {\mathbb{D}}_\alpha 
  \bar{\Phi}^{\bar{a}} \Big ] ~.
 \ee
 The superfields  $\Psi^\alpha_a$ are lagrange multipliers imposing the chirality constraint $ \bar{ \mathbb{D}}_{\pm} \Phi^{a} =0$.
 This can be written as
  \be
     S_{\rm dual}  = {\cal K} +  2 {\rm Re}( {\cal H})~,
   \ee
   where
    \be
    \label{calkis}
    {\cal K} =  \int d^2x~ d^4 \theta ~ K(\Phi^a, \bar{\Phi}^{\bar{a}})
   \ee
and
 \be
 \label{calhis}
 {\cal H} =  \int d^2x~ d^4 \theta ~
 \Psi^\alpha_a \bar{\mathbb{D}}_\alpha \Phi^a 
 =
  \int d^2x~ d^4 \theta ~( - i  \Psi_{-a}\bar{\mathbb{D}}_+ \Phi^a   + i \Psi_{+a} \bar{\mathbb{D}}_- \Phi^a   )~.
 \ee
 Note that the fields $\Psi^\alpha_a$ transform as (1,0) forms under holomorphic coordinate transformations for $M$ so that this action is covariant.

 We now return  to the geometrical setup.  We can enlarge the  K\"ahler target space $M$ with the coordinates $(z^a, \bar{z}^{\bar{a}})$
  by introducing odd additional coordinates $(\psi_{+a}, \bar{\psi}_{+\bar{a}})$ and $(\psi_{-a} , \bar{\psi}_{-\bar{a}})$  for
   two copies of the fibre of the odd cotangent bundle, so that we have
    $ (\Pi T^* \oplus  \Pi T^* )M$. (Here we follow the standard notation in which
    $\Pi$ indicates the parity-reversed fibre.)  Next we introduce the cotangent bundle to this supermanifold
    \be
      T^* \Big (  (\Pi T^* \oplus  \Pi T^* )M  \Big )
    \ee
     and  denote new fibre coordinates by $(p_a, \psi_{+}^a, \psi_{-}^a)$ (together with their complex conjugates).  This supermanifold
      is equipped with the canonical holomorphic symplectic structure 
      \be\label{new-sympl-hol}
  \omega'^{(2,0)} =    d p_a  \wedge d z^a  - i d \psi^{a}_+ \wedge d \psi_{-a} 
    +  i d \psi^{a}_- \wedge d \psi_{+a}  ~.
\ee
   Here the odd coordinates  $\psi_{\pm}^a$ transform as fibre coordinates for the
    odd tangent bundle. The transformations of $p_a$ are a bit more 
    complicated, the  $p_a$ transform as section of $T^*M$ plus an additional term which is quadratic in $\psi$. This supermanifold has 
     been studied explicitly and the detailed formulae for the transformation of all coordinates, together with other properties, can be found   in \cite{roytenberg1999courant} (e.g.\ see Remark 3.3.2  in \cite{roytenberg1999courant} for explicit formulae in the real case).  Next we   deform this 
      contangent bundle in the same way as described in subsection \ref{ss:global-kah} using the identification (\ref{KG-momgl})
       for the even coordinate $p_a$.  Thus we obtain the deformed cotangent bundle which is now the supermanifold ${\cal Z}'$
        and is equipped with the holomorphic symplectic structure given again by  formula (\ref{new-sympl-hol}) (in Darboux coordinates).

We  now extend the discussion of the previous subsection. Instead of using   the target space ${\cal Z}$ we    
work with the  supermanifold ${\cal Z}'$ .
 We  introduce the infinite dimensional space of supermaps
  from $\mathbb{R}^{2|4}$ to ${\cal Z}'$
  \be
   {\cal M} '= \{  \mathbb{R}^{2|4}\longrightarrow {\cal Z}' \}
  \ee
  with superfields $(\Phi^a, \Phi_a, \Psi^a_{\pm},  \Psi_{\pm a})$ (plus their complex conjugates). 
   Here  the superfields $\Phi^a$ are associated with coordinates $z^a$, the superfields $\Phi_a$ are associated with $p_a$, the superfields
    $\Psi_{\pm}^a$ are associated with $\psi_{\pm}^a$ and finally the superfields $\Psi_{\pm a}$ are associated with $\psi_{\pm a}$. The transformation rules for 
      the superfields  follow  from those of the coordinates of the  supermanifold ${\cal Z'}$.

  The space  $  {\cal M} '$ is equipped with
 the
  holomorphic symplectic form  
 \be
  \Omega_{\rm large} =  \int d^2x~ d^4 \theta ~ \Big ( \delta \Phi_a  \wedge \delta \Phi^a  - i \delta \Psi^{a}_+ \wedge \delta \Psi_{-a} 
    +  i \delta \Psi^{a}_- \wedge \delta \Psi_{+a}  \Big )~. 
\ee
 (Note that this formalism with an enlarged phase space has some analogies to the  BV/BRST formalism.) 
   We now use generating functions to define Lagrangian submanifolds.  We   define a holomorphic 
    Lagrangian submanifold of ${\cal Z}'$
     through the holomorphic generating function 
    of   $(\Phi^a, \Psi_{+a}, \Psi_{-a})$ given by $ {\cal H} $ defined in
    (\ref{calhis})
   which gives
    \be\label{holLa-Kah}
  \Phi_a = \frac{\delta {\cal H}}{\delta \Phi^a}=  i \bar{\mathbb{D}}_+ \Psi_{-a} - i \bar{\mathbb{D}}_- \Psi_{+a}~,~~~~\Psi_+^a = \frac{\delta {\cal H}}{\delta \Psi_{-a}}=\bar{\mathbb{D}}_+ \Phi^a~,~~~~
  \Psi_-^a = \frac{\delta {\cal H}}{\delta \Psi_{+a}}=      \bar{\mathbb{D}}_- \Phi^a~.
 \ee

         A real Lagrangian submanifold with respect to ${\rm Re}(\Omega_{\rm large})$ can be defined using the real generating function 
   of   $(\Phi^a, \Psi_{+a}, \Psi_{-a})$ given by ${\cal K} $ in  (\ref{calkis})
       so that we have
    \be\label{realLa-Kah}
      \Phi_a = \frac{\delta {\cal K}}{\delta \Phi^a} = \frac{\partial K}{\partial \Phi^a}~,~~~\Psi_+^a = \frac{\delta {\cal K}}{\delta \Psi_{-a}}=0~,
      ~~~~\Psi_-^a = \frac{\delta {\cal K}}{\delta \Psi_{+a}} = 0~.
    \ee
Remarkably, combining (\ref{holLa-Kah}) and (\ref{realLa-Kah}) and using
(\ref{sigpsi})
we arrive at the system of equations   (\ref{Kah-eqsmot}). 
 We interpret the equations of motion as the intersection of a  real 
 Lagrangian submanifold with a
  holomorphic Lagrangian submanifold. By enlarging the space from ${\cal M}$ to ${\cal M}'$
  we are able to encode both of these submanifolds in terms  of generating functions.

\section{Generalised K\"ahler geometry and sigma models}\label{s:general}

Next we   generalise the   discussion of the last section to the case of the general $N=(2,2)$ non-linear sigma model with target space a generalised K\"ahler manifold. 
These are manifolds equipped with two complex structures, $I_+$ and $I_-$,
and a metric that is hermitian with respect to both of them. They are also equipped with a closed 3-form $H$; see
  \cite{Gualtieri:2014kja} for details of the geometry.
 In this section we will focus on the local structure and  
  we postpone  the discussion of global geometrical issues until the next section. 

\subsection{The Generalised K\"ahler Sigma Model and its Dual Formulation}

The general $N=(2,2)$ non-linear sigma model can be written in terms of four types of
$N=(2,2)$ superfields \cite{Lindstrom:2005zr}.
These are the chiral and anti-chiral fields
 \be
 \bar{ \mathbb{D}}_{\pm} \phi^{a} =0~,~~~~~~
 \mathbb{D}_{\pm} \bar{\phi}^{\bar{a}} =0~,
 \ee
the twisted chiral  and twisted anti-chiral fields  
\be
 \bar{\mathbb{D}}_+ \chi^{a'}=0~,~~~\mathbb{D}_- \chi^{a'}=0~,~~~
 \mathbb{D}_+ \bar{\chi}^{\bar{a}'}=0~,~~~\bar{\mathbb{D}}_- \bar{\chi}^{\bar{a}'}=0~, 
\ee
the left semichiral fields 
\be
 \bar{\mathbb{D}}_+ X^{n}_L =0~,~~~ \mathbb{D}_+ \bar{X}_L^{\bar{n}'}=0~,
\ee
 and finally the right semichiral fields
\be
 \bar{\mathbb{D}}_- X_R^{n'}=0~,~~~ \mathbb{D}_- \bar{X}_R^{\bar{n'}}=0~. 
\ee
Each superfield has a bosonic component that is interpreted as  one of the coordinates of the target space, so the 
split into four kinds of superfields corresponds to a split of the coordinates into four different kinds. 
  This is always possible locally, and will be possible globally if we assume that all relevant Poisson structures are regular   \cite{Lindstrom:2005zr}; we shall assume that this is the case
   for the rest of the paper.
Then the coordinates are $\varphi ^A= (\phi^{a},\chi^{a'}, X^{n}_L,X_R^{n'})$ plus their complex conjugates and the 
 indices split  as $A = (a, a', n, n')$,
   $\bar{A} = (\bar{a}, \bar{a}', \bar{n}, \bar{n}')$.
 There are equal numbers of left semi-chirals  $X_L$ and right semi-chirals  $X_R$.
Then 
the general $N=(2,2)$ sigma model action is 
\be\label{GKG-action-general}
  S = \int d^2x~ d^4 \theta ~K(\phi, \bar{\phi}, \chi, \bar{\chi}, X_L, \bar{X}_L, X_R, \bar{X}_R) ~,
\ee
  where  $K$ is real function of the   superfields; see  \cite{Lindstrom:2005zr} for further explanation. 
     
   In analogy with the K\"ahler case (\ref{dual-Kah}) we can rewrite the action in first-order form.
  We replace the constrained superfields $\varphi ^A= (\phi^{a},\chi^{a'}, X^{n}_L,X_R^{n'})$ with unconstrained superfields
  $\Phi ^A= (\Phi ^{a},\Phi ^{a'}, \Phi ^{n}_L,\Phi _R^{n'})$ and introduce constrained lagrange multiplier fields 
  $\varphi _A=( \Sigma _a ,
   \Lambda_{a'} , Y_{L n} , \bar{Y}_{R\bar{n}'} )$ (plus their complex conjugates)
  whose field equations impose the appropriate constraints on the fields $\Phi ^A$.
  The action is
   \be\label{dual-GKs}
  S_{\rm dual} = \int d^2x~ d^4 \theta ~\Big ( K(\Phi, \bar{\Phi}) - \varphi _A\Phi ^A-
 \bar \varphi _{\bar A}\bar\Phi ^{\bar A} \Big )~,
  \ee
   which can be expanded to give
   \bea\label{dual-GK}
  S_{\rm dual} = \int d^2x~ d^4 \theta ~\Big ( K(\Phi, \bar{\Phi}) - \Sigma _a \Phi^a - \bar{\Sigma}_{\bar{a}} \bar{\Phi}^{\bar{a}} -
   \Lambda_{a'} \Phi^{a'} - \bar{\Lambda}_{\bar{a}'} \bar{\Phi}^{\bar{a}'}
   \nonumber
   \\
    - Y_{L n} \Phi^n - \bar{Y}_{L\bar{n}} \bar{\Phi}^{\bar{n}}
    - Y_{Rn'} \Phi^{n'} -  \bar{Y}_{R\bar{n}'} \bar{\Phi}^{\bar{n}'} \Big )~. 
\eea
As seen in the last section, the $\Sigma_a$ are complex linear superfields. The constraints on the other fields are as follows:
the $\Lambda_{a'}$ are twisted complex
  linear superfields, the $Y_{Ln}$ are left semichiral superfields and the $Y_{Rn'}$ are right semichiral superfields (see the appendix for the definitions). 
   The field equations for  $\Sigma$, $\Lambda$, $Y_L$ and $Y_R$ impose the appropriate chirality constraints on the $\Phi$'s: they impose that $\Phi ^{a}$ is chiral, $\Phi ^{a'}$ is twisted chiral, $\Phi ^{n}_L$ is left semi-chiral and $\Phi _R^{n'}$ is right semi-chiral.
 Then integrating out  the fields $\varphi _A$ we recover
  the original action (\ref{GKG-action-general})
  (after renaming the fields $\Phi ^A= (\Phi ^{a},\Phi ^{a'}, \Phi ^{n}_L,\Phi _R^{n'})$ as $\varphi _A=( \Sigma _a ,
   \Lambda_{a'} , Y_{L n} \Phi^n , \bar{Y}_{R\bar{n}'} )$).

    Alternatively we can integrate out the $\Phi$'s to arrive at the dual action
    \be\label{dual-GKa}
  S_{\rm dual} = \int d^2x~ d^4 \theta ~\tilde{K} (\Sigma, \bar{\Sigma}, \Lambda, \bar{\Lambda}, Y_L, \bar{Y}_L, Y_R, \bar{Y}_R)~,
\ee
where
   $\tilde{K} (\Sigma, \bar{\Sigma}, \Lambda, \bar{\Lambda}, Y_L, \bar{Y}_L, Y_R, \bar{Y}_R)$   is a generalised Legendre 
    transform of $K$.\footnote{This duality and the action (\ref{dual-GK}) have also been considered by U. Lindstr\" om, in unpublished work.}
  As in the last section, the duality needs  the Legendre transform to be well-defined, which requires $K$ to be convex.

\subsection{Lagrangian System for the Generalised K\"ahler Sigma Model}

The field equations that follow from (\ref{dual-GK}) can be re-expressed as the following system of equations for constrained superfields
$\varphi ^A= (\phi^{a},\chi^{a'}, X^{n}_L,X_R^{n'})$ and $\varphi _A=( \Sigma _a ,
   \Lambda_{a'} , Y_{L n}  , \bar{Y}_{R\bar{n}'} )$:  
 \be 
   \varphi _A= \frac{\partial K}{\partial \varphi ^A}~,
   \ee
where $K(\varphi ^A, \bar \varphi ^{\bar A})$ is the generalised K\"ahler potential.
Explicitly, this gives the set of equations

\fbox{
 \addtolength{\linewidth}{-2\fboxsep}%
 \addtolength{\linewidth}{-2\fboxrule}%
 \begin{minipage}{\linewidth}
\bea\label{general-eqmot}
 && \Sigma_a = \frac{\partial K}{\partial \phi^a}~,~~~~\Lambda_{a'} =  \frac{\partial K}{\partial \chi^{a'}}~,~~~~
    Y_{Ln}=  \frac{\partial K}{\partial X^n_L}~,~~~~Y_{Rn'} =  \frac{\partial K}{\partial X^{n'}_R} \nonumber \\ 
 &&    \bar{\mathbb{D}}_+ \bar{\mathbb{D}}_- \Sigma _a=0~,~~~\bar{\mathbb{D}}_+ \mathbb{D}_- \Lambda_{a'}=0~,~~~\bar{\mathbb{D}}_+ Y_{Ln}=0~,~~~~\bar{\mathbb{D}}_- Y_{Rn'}=0 \\
  &&   \bar{\mathbb{D}}_{\pm} \phi^a=0~,~~~\bar{\mathbb{D}}_+ \chi^{a'} = \mathbb{D}_- \chi^{a'} =0~,~~~\bar{\mathbb{D}}_+ X^n_L=0~,~~~~\bar{\mathbb{D}}_- X^{n'}_R=0 \nonumber
\eea
 \end{minipage}
}
~\\
 together with the complex conjugate equations.  
 Regarding these as equations for the constrained superfields $ \varphi ^A$,
 the constraints on $\varphi _A$ applied to $\frac{\partial K}{\partial \varphi ^A}$ give the field equations
for  $ \varphi ^A$ that follow from the action (\ref{GKG-action-general}).

If a Legendre transform $\tilde K$ exists, then the equations can also be cast in the dual form
\be 
 \varphi ^A  = \frac{\partial \tilde K}{\partial \varphi _A }~,
   \ee
where $\tilde K(\varphi _A, \bar \varphi _{\bar A})$ is the Legendre transform of the generalised K\"ahler potential.
Explicitly, this gives the set of equations

\fbox{
 \addtolength{\linewidth}{-2\fboxsep}%
 \addtolength{\linewidth}{-2\fboxrule}%
 \begin{minipage}{\linewidth}
\bea\label{general-eqmot-dual}
&& \phi^a  = \frac{\partial \tilde K}{\partial\Sigma_a }~,~~~~
 \chi^{a'}=  \frac{\partial \tilde K}{\partial \Lambda_{a'} }~,~~~~
   X^n_L =  \frac{\partial  \tilde K}{\partial Y_{Ln}  }~,~~~~
    X^{n'}_R =  \frac{\partial \tilde K}{\partial  Y_{Rn'} } 
   \nonumber \\ 
&&     \bar{\mathbb{D}}_+ \bar{\mathbb{D}}_- \Sigma _a=0~,~~~\bar{\mathbb{D}}_+ \mathbb{D}_- \Lambda_{a'}=0~,~~~\bar{\mathbb{D}}_+ Y_{Ln}=0~,~~~~\bar{\mathbb{D}}_- Y_{Rn'}=0 \\
 &&    \bar{\mathbb{D}}_{\pm} \phi^a=0~,~~~\bar{\mathbb{D}}_+ \chi^{a'} = \mathbb{D}_- \chi^{a'} =0~,~~~\bar{\mathbb{D}}_+ X^n_L=0~,~~~~\bar{\mathbb{D}}_- X^{n'}_R=0 \nonumber
\eea
 \end{minipage}
}
~\\
 together with the complex conjugate equations.  We leave this dual system for  future discussion and concentrate here  on the equations 
  (\ref{general-eqmot}).
 
 \subsection{Local symplectic interpretation}\label{ss:local-sympl}
 
 We now take a first look at the symplectic geometry associated with the system (\ref{general-eqmot}). 
 We will  investigate the local geometry in this subsection
 and 
   postpone discussion of the global structure to the following section.
  In particular, we restrict ourselves  to a single coordinate patch  $U$
  with coordinates $ \varphi ^A, \varphi _A$. This coordinate patch  has a  complex structure with respect to which
  $(\varphi ^A, \varphi _A)=(\phi^a, \Sigma_a, \chi^{a'}, \Lambda_{a'}, 
   X_L^n, Y_{L n}, X_R^{n'}, Y_{Rn'})$ are holomorphic coordinates and their complex conjugates are anti-holomorphic coordinates.
  
 Following our discussion  of the K\"ahler case, we re-cast the theory in terms of unconstrained superfields,
 replacing the  constrained superfields
 $ \varphi ^A, \varphi _A$ with unrestricted fields $\Phi^A, \Phi_A$ 
 (together with their complex conjugates $\bar{\Phi}^{\bar{A}},\bar{\Phi}_{\bar{A}}$).
 First, we regard $\Phi^A, \Phi_A$ as coordinates on $U$.
 Then $U$ has a holomorphic symplectic structure
  \be
  \omega^{(2,0)} = d \Phi_A  \wedge d \Phi^A ~.
  \ee
 
 Following our discussion of the K\"ahler case, we now consider unrestricted superfields $\Phi^A(x,\theta,\bar \theta), \Phi_A(x,\theta,\bar \theta)$ which map superspace to the set $U$.
 Of course, we will be interested in the generalisation to superfields mapping to the whole space rather than to the subset $U$, but for now we work with this restricted case.
 Then the symplectic form $\omega^{(2,0)}$ lifts to 
 \be
  \Omega =  \int d^2x~ d^4 \theta ~   \delta \Phi_A  \wedge \delta \Phi^A ~.
  \ee
     As in the K\"ahler case, the first line in equation (\ref{general-eqmot}) corresponds to a  submanifold of the space of supermaps to $U$
   \be
    \Phi_A = \frac{\partial K}{\partial \Phi^A}
   \ee
  that is real Lagrangian with respect to  ${\rm Re}(\Omega)$. The superfield constraints in  (\ref{general-eqmot}) define a holomorphic isotropic submanifold, as $\Omega$ vanishes when restricted to superfields satisfying the constraints: 
   \be
    \Omega \Big| _{\rm constraints} =   \int d^2x~ d^4 \theta ~  \Big (  \delta \Phi_a  \wedge \delta \Phi^a +  \delta \Phi_{a'}  \wedge \delta \Phi^{a'}
    +  \delta \Phi_n  \wedge \delta \Phi^n + \delta \Phi_{n'}  \wedge \delta \Phi^{n'} \Big  ) =0~,
   \ee
   where we have used the superfield representations (\ref{reps-super}). Thus we see that the corresponding submanifold is holomorphic 
    isotropic with respect to the  complex structure on $U$. Using the language of generating functions we now
     show that the corresponding manifold arises as  a Lagrangian submanifold of an enlarged space.  
   
As in the K\"ahler case, we enlarge our space 
 by adding additional fermionic fields $\Psi_{+A}, \Psi_{-A}$
 and take the symplectic form to be
 \be
  \Omega_{\rm large} =  \int d^2x~ d^4 \theta ~ \Big ( \delta \Phi_A  \wedge \delta \Phi^A  - i \delta \Psi^{A}_+ \wedge \delta \Psi_{-A} 
    +  i \delta \Psi^{A}_- \wedge \delta \Psi_{+A}  \Big )~.
\ee
 Next we choose two generating functions which depend on
 $(\Phi^A, \Psi_{+A}, \Psi_{-A})$. We define a real generating function
\be
    {\cal K} =  \int d^2x~ d^4 \theta ~ K(\Phi^A, \bar{\Phi}^{\bar{A}})~,
   \ee
    which gives rise to a real Lagrangian submanifold with respect to  ${\rm Re}(\Omega)$. We also define the
   holomorphic generating function 
 \be
 {\cal H} =   \int d^2x~ d^4 \theta ~ \Big (  i  \Psi_{-a}\bar{\mathbb{D}}_+ \Phi^a   - i \Psi_{+a} \bar{\mathbb{D}}_- \Phi^a   +
 i  \Psi_{-a'}\bar{\mathbb{D}}_+ \Phi^{a'}   - i \Psi_{+a'} \mathbb{D}_- \Phi^{a'}  +
  i  \Psi_{-n}\bar{\mathbb{D}}_+ \Phi^n   - i \Psi_{+n'} \bar{\mathbb{D}}_- \Phi^{n'} \Big  )~,
 \ee
 which defines a holomorphic Lagrangian submanifold. The intersection of these two Lagrangian submanifolds give rise to the equations (\ref{general-eqmot}). 
 
 However, this local discussion, attractive though it is, does not in general extend to the full space as the  local complex structure on $U$
 used here does not in general extend to a complex structure on the whole space, because the transition functions mix coordinates that are holomorphic with respect to the  local complex structure with ones that are anti-holomorphic.

\section{Global issues}\label{s:examples}

In this section, we seek  a global structure that is consistent with the equations (\ref{general-eqmot}).
Recall that in the K\"ahler case we considered the sigma model's symmetries -- the holomorphic diffeopmorphisms and the K\"ahler gauge transformations -- and saw in section \ref{s:kahler} how requiring these to be symmetries of the equations (\ref{Kah-eqsmot}) fixed the transformation of $\Sigma$ under these transformations. This then led to a global structure in which these symmetries were used in the glueing relations.
In this section, we will
 study the symmetries of the equations (\ref{general-eqmot}) 
 and attempt to define a global structure  by using these symmetries as glueing relations in the overlaps of patches.

 There are two  kinds of symmetry here. The first consists of  those diffeomorphisms 
 $\Phi^A\to {\Phi '}^A(\Phi^A,\bar\Phi^A)$
 that
  are  compatible
   with the choice of coordinates for symplectic foliations and with the various superfield constraints; these transformations are given in   (\ref{superf-transf}).
  Requiring these to extend to symmetries of (\ref{general-eqmot}) then determines
  the transformations of the fields $\varphi_A$, and these are then used in the glueing relations.
   The second   kind of symmetry arises from  the fact that the generalised K\"ahler 
    potential is not uniquely defined, but is only defined  up to transformations that generalise (\ref{Kshift}).
    On the intersection of two patches $U_\alpha \cap U_\beta$ the generalised K\"ahler potentials are  related by  the  transformations \cite{Hull:2008vw}
\be\label{GKP-ambig}
    K_\alpha - K_\beta = F_{\alpha\beta}^+ (\phi, \chi, X_L) + \bar{F}^+(\bar{\phi}, \bar{\chi}, \bar{X}_L) + 
     F_{\alpha\beta}^- (\phi, \bar{\chi}, X_R) + \bar{F}_{\alpha\beta}^-(\bar{\phi}, \chi, \bar{X}_R)~,
   \ee 
   since the combinations on the right hand side vanish identically after superintegration. 
  For this to be a symmetry of the system of equations    (\ref{general-eqmot})  we require that
  the   fields $\varphi_A$ are glued  on the intersection of two patches $U_\alpha \cap U_\beta$ as follows
\bea
&&    ( \Sigma_a)_\beta  =
( \Sigma_a )_\alpha + \frac{\partial}{\partial \phi^a} \Big ( F_{\alpha\beta}^+ (\phi, \chi, X_L)  +   F_{\alpha\beta}^- (\phi, \bar{\chi}, X_R)  \Big )~, \label{gen-shift-1} 
\\
  &&    ( \Lambda_{a'} )_\beta  =
(  \Lambda_{a'} )_\alpha 
+ \frac{\partial}{\partial \chi^{a'}} \Big ( F_{\alpha\beta}^+ (\phi, \chi, X_L)  + \bar{F}_{\alpha\beta}^-(\bar{\phi}, \chi, \bar{X}_R) \Big )~,  \label{gen-shift-2} \\ 
  &&  (Y_{Ln})_\beta  =
(  Y_{Ln} )_\alpha 
+  \frac{\partial}{\partial X^n_L} F_{\alpha\beta}^+ (\phi, \chi, X_L) ~,  \label{gen-shift-3} \\
   && ( Y_{R n'} )_\beta  =
(  Y_{R n'})_\alpha 
+  \frac{\partial}{\partial X^{n'}_R} F_{\alpha\beta}^- (\phi, \bar{\chi}, X_R) ~, \label{gen-shift-4}
    \eea
   (These glueing relations are to be composed with the transition functions expressing the change of coordinates between the two patches.)
    As explained in   Appendix B, the above shifts satisfy the cocycle conditions on the triple intersections $U_\alpha\cap U_\beta\cap U_\gamma$.
    
    Both kinds of   symmetry should be compatible with the superfield 
     redefinitions (\ref{superf-transf}) allowed by the chirality constraints.  It will be useful to state the most 
      general superfield redefinitions 
       that are compatible with the superfield constraints
       (we suppress all indices for the sake of clarity) 
        \bea
&& \phi~\longrightarrow~f_1(\phi)~, \nn \\
&& \Sigma~\longrightarrow~f_2(\phi)\Sigma  + f_3 (\phi, \chi, X_L, Y_L) + f_4(\phi, \bar{\chi}, X_R, Y_R) ~,\nn \\
 &&\chi~\longrightarrow~f_5(\chi)~,\label{superf-transf-double} \\
 && \Lambda ~\longrightarrow~ f_6(\chi) \Lambda + f_7 (\phi, \chi, X_L, Y_L) + f_8 (\bar{\phi}, \chi, \bar{X}_R, \bar{Y}_R)  ~,\nn \\
 && X_L ~\longrightarrow~ f_9(\phi, \chi, X_L, Y_L)~, \nn \\
&& X_R~\longrightarrow~f_{10}(\phi, \bar{\chi}, X_R, Y_R)~,\nn \\
&& Y_L ~\longrightarrow~ f_9(\phi, \chi, X_L, Y_L)~, \nn \\
&& Y_R~\longrightarrow~f_{10}(\phi, \bar{\chi}, X_R, Y_R)~,\nn
 \eea
  where $f_i$ are arbitrary functions of their arguments.  
  We will use these field redefinitions as a guiding principle in our discussion
   of the global issues.

\subsection{Sigma models with  semi-chiral fields only}\label{ss:only-semis}

We consider first the special case in which the generalised K\"ahler geometry is described in terms of semi-chiral fields only. 
The system of equations (\ref{general-eqmot}) 
for the case 
 with only semi-chiral fields reduces to
  
  \fbox{
 \addtolength{\linewidth}{-2\fboxsep}%
 \addtolength{\linewidth}{-2\fboxrule}%
 \begin{minipage}{\linewidth}
\bea\label{semi-eqmot}
    Y_{Ln}=  \frac{\partial K}{\partial X^n_L}~,~~~~Y_{Rn'} =  \frac{\partial K}{\partial X^{n'}_R} \nonumber \\ 
    \bar{\mathbb{D}}_+ Y_{Ln}=0~,~~~~\bar{\mathbb{D}}_- Y_{Rn'}=0 \\
     \bar{\mathbb{D}}_+ X^n_L=0~,~~~~\bar{\mathbb{D}}_- X^{n'}_R=0 \nonumber
\eea
 \end{minipage}
}
~\\
We now look at the 
symmetries of this system.
 The constraints on the superfields allow
 the following transformations 
 \bea
 \label{xytrans}
 & X_L^n\rightarrow f^n(X_L, Y_L)~,~~~Y_{L n}\rightarrow g_{n} (X_L, Y_L)~,\\
  & X_R^{n'}\rightarrow h^{n'}(X_R, Y_R)~,~~~Y_{Rn'}\rightarrow k_{n'} (X_R, Y_R)~,
 \eea
 for arbitrary functions $f,g,h,k$
  and  the shifts (\ref{gen-shift-3})-(\ref{gen-shift-4}) are just a special form of these transformations. 
  This suggests that 
   there is no linear structure associated with the $Y$ directions so that, unlike the K\"ahler case, the conjugate coordinates are not fibre coordinates for some vector bundle.

    The correct global interpretation of the system  (\ref{semi-eqmot})     follows from \cite{Lindstrom:2005zr}.
    We now briefly review the relevant geometry.
     We introduce local coordinates $Z^A = (\zeta_L ^n,  \zeta^{n'}_R)$ on the manifold $M$ corresponding to the lowest components of the superfields $\varphi^A = (X_L^{n}, X_R^{n'})$
        together with variables $P_A = (\pi_{Ln},  \pi_{Rn'})$ corresponding to the lowest components of the superfields $\varphi_A = (Y_{Ln}, Y_{Rn'})$. 
        The space $M$ has two complex structures $I_\pm$ and it was shown in \cite{Lindstrom:2005zr} that $(\zeta_L^n, \pi_{Ln})$ can   be taken as coordinates for $M$ and that moreover they are holomorphic coordinates with respect to $I_+$. The
         symplectic form 
         $ d\pi_{Ln} \wedge d\zeta^n_L$ is then holomorphic with respect to $I_+$. 
        Similarly,
        $(\zeta^{n'}_R, \pi_{Rn'})$ can    be also taken as coordinates for $M$ and these 
        are holomorphic coordinates with respect to $I_-$.   The symplectic structure $  d\pi_{Rn'} \wedge d\zeta^{n'}_R$  is
        holomorphic  with respect to $I_-$. The transformation from the coordinates $(\zeta_L, \pi_L)$ to the coordinates $(\zeta_R, \pi_R)$ is a canonical transformation that is generated by the generalised K\"ahler potential $K(\zeta_L, \zeta_R)$.
  
  This extends to a global formulation. In each patch $U$ for $M$, there are $I_+$-holomorphic coordinates $(\zeta_L, \pi_L)$ and in each overlap $U\cap U'$  the coordinates $(\zeta_L, \pi_L)$ in $U$ and the coordinates $(\zeta'_L, \pi'_L)$ in $U'$ are related by     
 $I_+$-holomorphic reparameterisations, $\zeta'_L=\zeta'_L(\zeta_L,\pi_L)$, $\pi_L'=\pi'_L(\zeta_L,\pi_L)$. Similarly, $I_-$-holomorphic
  coordinates $(\zeta^{n'}_R, \tilde\pi_{Rn'})$ can also be introduced for each patch
and  the transition functions for the coordinates $(\zeta_R, \pi_R)$ are $I_-$-holomorphic reparameterisations $\zeta'_R=\zeta_R'(\zeta_R,\pi_R)$, $\pi'_R=\pi_R'(\zeta_R,\pi_R)$.
 These transition functions give glueing relations for the corresponding superfields that
 are  consistent with (\ref{xytrans}), so that they preserve the constraints.
 Then in each patch, there is a generalised K\"ahler potential $K(\zeta_L, \zeta_R)$ generating the transformation between the $I_+$-holomorphic coordinates $(\zeta_L^n, \pi_{Ln})$ and the $I_-$-holomorphic
 coordinates $(\zeta^{n'}_R, \pi_{Rn'})$ and the potentials in overlapping patches have the glueing relations (\ref{GKP-ambig}).
        
 This can be elegantly reformulated in terms of a double space ${\cal Z}= M\times M$ with coordinates $(\zeta_L, \pi_L, \zeta_R,  \pi_R)$.
 The first $M$ is taken to have coordinates $(\zeta_L, \pi_L)$ which are    holomorphic with respect to $I_+$. It has the
  symplectic form 
         \be
         \omega^{(2,0)}_+= d\pi_{Ln} \wedge d\zeta^n_L
         \ee
          which is  holomorphic with respect to $I_+$. The second
         $M$ is taken to have coordinates $(\zeta_R, \pi_R)$ that are holomorphic with respect to $I_-$ and an $I_-$-holomorphic symplectic form 
         \be
         \omega^{(2,0)}_-=   d\pi_{Rn'} \wedge d\zeta^{n'}_R
         ~.
         \ee 
  The space ${\cal Z}= M\times M$ is itself equipped with the 
  complex structure
  \be I=I_+\times I_-
  \ee and the
   symplectic structure 
           \be
            \omega^{(2,0)} = dP_A \wedge dZ^A = 
            d\pi_{Ln} \wedge d\zeta^n_L+d\pi_{Rn'} \wedge d\zeta^{n'}_R~,
           \ee
      which is holomorphic with respect to $I$.
        The splitting of the coordinates into $P$ and $Z$ is a choice of polarization and $P,Z$ are Darboux coordinates.  
 It will be useful to write this as
        \be
        {\cal Z}= M_+\times M_-~,
        \ee
        where $M_+=(M,I_+)$ and $M_-=(M,I_-)$.
        
 Then the Lagrangian submanifold of        ${\cal Z}$ defined by
 the equation  with a real function $K(\zeta_L ^n, \bar{\zeta}_L^{\bar{n}},   \zeta^{n'}_R, \bar{\zeta}_R^{\bar{n}'})$
  \be 
    P_A = \frac{\partial K}{\partial Z^A}
  \ee
  gives a diagonally embedded submanifold $M$, with coordinates 
  $(\zeta_L ^n,  \zeta^{n'}_R)$, and this is the sigma model target space.
  
  The space 
  ${\cal Z}= M_+\times M_-$ is then the setting for the equations (\ref{semi-eqmot}). Then $X_L,Y_L$ are supermaps to the first factor $M_+$ and $X_R,Y_R$ are supermaps to the second factor $M_-$  while $X_L,X_R$ are supermaps to the Lagrangian submanifold $M$.

We now reformulate this in terms of \emph{unconstrained} superfields 
$\Phi_A  , \Phi^A$ where $ \Phi^A= (\zeta_L(x,\theta, \bar \theta), \zeta_R(x,\theta, \bar \theta))$ and $\Phi_A=(\pi_L(x,\theta, \bar \theta), \pi_R(x,\theta, \bar \theta))$.
The
space of supermaps  
\be
   {\cal M} = \{ \Phi :\mathbb{R}^{2|4}\longrightarrow M_+\times M_- \}
  \ee
 is equipped with the holomorphic symplectic structure 
  \be
  \Omega =  \int d^2x~ d^4 \theta ~   \delta \Phi_A  \wedge \delta \Phi^A~. 
  \ee
    The equations of motion (\ref{semi-eqmot}) then specify the intersection of 
    two 
    submanifolds of $ {\cal M}$.
    The submanifold determined by
      \be
    \Phi_A = \frac{\partial K}{\partial \Phi^A}
   \ee
    is a real Lagrangian submanifold with respect to ${\rm Re} (\Omega)$, while the
    submanifold specified by the semi-chiral superfield constraints is a  holomorphic isotropic submanifold with respect to $\Omega$.
    As before, enlarging the space of supermaps by introducing auxiliary fermionic superfields 
    gives a formulation in which the 
  submanifold (of the enlarged space) specified by the semi-chiral superfield constraints   
    is Lagrangian.
   
\subsection{Sigma models with  semi-chiral fields and chiral fields}
\label{semichichi}
   
   For our next example, we consider  the case in which there are chiral and semichiral superfields but there are no $(\chi, \Lambda)$-fields 
   in the equations   (\ref{general-eqmot}). As we shall see, this combines features of the K\"ahler case (with only chiral fields) with features of the purely semi-chiral case from 
    the previous subsection.  The most general field redefinitions  (\ref{superf-transf}) compatible with the constraints 
  can be written in this case as follows
  \bea
&& \phi~\longrightarrow~f_1(\phi)~, \nn \\
&& \Sigma~\longrightarrow~f_2(\phi)\Sigma  + f_3 (\phi,  X_L, Y_L) + f_4(\phi,  X_R, Y_R) ~,\nn \\
 && X_L ~\longrightarrow~ f_9(\phi,  X_L, Y_L)~,~~~ Y_L ~\longrightarrow~ \tilde{f}_9(\phi,  X_L, Y_L)~, \nn \\
&& X_R~\longrightarrow~f_{10}(\phi, X_R, Y_R)~,~~~Y_R~\longrightarrow~\tilde{f}_{10}(\phi, X_R, Y_R)~,
\label{trsa}
 \eea
 (together with their complex conjugates) for some  functions $f_i$. 
  Importantly,  there is no mixing between barred and unbarred 
  fields in this case.  Thus the holomorphic superfields $(\phi, \Sigma, X_L, Y_L, X_R, Y_R)$  are glued to each other and the transition functions do not depend on the conjugate
   fields and thus we have a complex structure on the space of superfields.  Here the glueing combines two models, the deformed holomorphic cotangent bundle 
    from the  K\"ahler case for the  $(\phi, \Sigma)$ coordinates and the product $M\times M$ for the semi-chiral coordinates. 
    
 We now explain the construction in more detail.
 We introduce local coordinates $Z^A = (z^a,\zeta _L^n,  \zeta^{n'}_R)$ on the manifold $M$ corresponding to the lowest components of the superfields $\varphi^A = (\phi^a,X_L^{n}, X_R^{n'})$
        together with variables $P_A = (p_a,\pi_{Ln}, \pi_{Rn'})$ corresponding to the lowest components of the superfields $\varphi_A = (\Sigma_a, Y_{Ln}, Y_{Rn'})$. 
    The  glueing relations (\ref{GKP-ambig}) for the 
        generalised K\"ahler potential $K(z,\bar z,\zeta_L,\zeta_R,\bar\zeta_L,\bar\zeta_R)$
       in an overlap
        $U\cap U'$  become in this case
        \be K'-K=F^+(z,\zeta_L)+ F^-(z,\zeta_R)+ {\rm complex ~conjugate}
        \label{Ktrans}
        \ee
        where $K$ is the potential on $U$ and $K'$ is the potential on $U'$.
     With $Z^A$ the coordinates of $M$, the $P_A$ can be taken to be defined by
     \be
P_A =   \frac {\partial K} {\partial Z^A}
\ee 
Then (\ref{Ktrans}) implies that the glueing conditions for $P_A$ in an overlap
        $U\cap U'$ must be
   \be\label{p-glue-stype}
        p'_a=p_a + \frac {\partial} {\partial z^a}
        \left[ F^+(z,\zeta_L)+ F^-(z,\zeta_R) \right]~,
        \ee
      \bea
       && \pi'_{Ln} = \pi_{Ln} + \frac {\partial} {\partial \zeta_L^n} F^+(z,\zeta_L) ~,
       \nonumber\\
       && \pi'_{Rn'} = \pi_{Rn'} + \frac {\partial} {\partial \zeta_R^{n'}} F^-(z,\zeta_R)~,
       \label{piglue}
      \eea
where $F^\pm$ are the functions appearing in (\ref{Ktrans}) and these glueing relations satisfy the cocycle condition in triple overlaps (see Appendix B). 
Note that (\ref{piglue})
are a special case of the transformations (\ref{transf-sympt}).
Then the variables $(Z^A,P_A)$ can be regarded as coordinates  on a manifold, which we refer to as the
\lq phase space'. We construct this phase space ${\cal Z}$ explicitly below.

On $M$ we can use the coordinates $Z^A = (z^a,\zeta _L^n,  \zeta^{n'}_R)$, or the $I_+$-holomorphic coordinates $ (z
,\zeta _L ,  \pi _L)$ or the $I_-$-holomorphic coordinates $ (z
,\zeta _R ,  \pi _R)$.
        The manifold $M$ is foliated by subspaces of constant $z$. The coordinates  on each leaf can be taken to be $\zeta _L,  \zeta_R$ or
        $\rho_L= (\zeta_L,\pi_L)$ or $\rho_R= ( \zeta_R,\pi_R)$.
       Choosing the leaf coordinates to be  $\rho_L= (\zeta_L,\pi_L)$, 
        the transition functions in an overlap
        $U\cap U'$ are $I_+$-holomorphic and of the form
        \be
        z'=z'(z)~,\qquad \rho_L' =\rho_L'(z,\rho_L)~,
        \ee
        which is   consistent with (\ref{trsa}).
        Alternatively, choosing
        the coordinates  on each leaf to be  $\rho_R= ( \zeta_R,\pi_R)$,
        the transition functions in an overlap
        $U\cap U'$ are $I_-$-holomorphic of the form
        \be
        z'=z'(z)~,\qquad \rho_R' =\rho_R'(z,\rho_R)~.
        \ee 
               
    We now construct a space   $\hat M$ in which the leaves of this foliation are \lq doubled'.
 For each open set $U$ of an atlas for $M$ we can choose coordinates $(z,\rho_L)$ or
        $(z,\rho_R)$ adapted to the foliation. Then for each  such $U$ we introduce a space
        $\hat U$ with coordinates $z,\rho_L,\rho_R$. Then $\hat M$ is constructed by glueing together the patches $\hat U$ with transition functions
        \be\label{transf-sympt} 
z'=z'(z)~,\qquad \rho_L' =\rho_L'(z,\rho_L)~,\qquad \rho_R' =\rho_R'(z,\rho_R)~.
        \ee
        Note that the foliated structure of $M$ is essential for this construction of $\hat M$.
        This space $\hat M$ can be regarded as a quotient of $M_+\times M_-$.
        With coordinates $(z,\rho_L)$ for $M_+$ and
        $(z',\rho_R)$ for $M_-$, taking the quotient by the relation $z\sim z'$ gives $\hat M$. 
        
        The next step is to construct a bundle ${\cal Z}$ over $\hat M$ with fibre coordinates $p_a$ and
        glueing conditions
        (\ref{p-glue-stype}).
        It is important that these satisfy the cocycle condition in triple overlaps.
              Then ${\cal Z}$  has coordinates $Z^A,P_A$ and the transition functions are all holomorphic in $Z^A,P_A$ and so this endows ${\cal Z}$ with a complex structure. Moreover, ${\cal Z}$ has a holomorphic symplectic structure
        \be
        \omega^{(2,0)}=
        dP_A\wedge dZ^A=dp_a\wedge dz^a + d\pi_{Ln}\wedge d \zeta _L^n
        + d\pi_{Rn'}
        \wedge d \zeta^{n'}_R~,
        \ee
      which is invariant under the glueing (\ref{p-glue-stype})  supplemented by the diffeormorphisms 
      (\ref{piglue}).
The equation
\be
P_A =   \frac {\partial K} {\partial Z^A}
\ee 
specifies a real Lagrangian submanifold of ${\cal Z}$
with respect to ${\rm Re}(\omega^{(2,0)})$, which is the original manifold $M$. 

We now consider the space of unrestricted supermaps $\Phi^A=Z^A(x,\theta,\bar\theta)$,
$\Phi_A=P_A(x,\theta,\bar\theta)$ 
\be
   {\cal M} = \{ \Phi :\mathbb{R}^{2|4}\longrightarrow {\cal Z} \}~,
  \ee
 which has the holomorphic symplectic structure 
  \be
  \Omega =  \int d^2x~ d^4 \theta ~   \delta \Phi_A  \wedge \delta \Phi^A~. 
  \ee 
  The submanifold determined by
      \be
    \Phi_A = \frac{\partial K}{\partial \Phi^A}
   \ee
    is a real Lagrangian submanifold with respect to ${\rm Re} (\Omega)$
    and is the space of unconstrained supermaps to $M$.
    The superfield
    constraints that $z^a(x,\theta,\bar\theta)=\phi^a$ is chiral,
    $p_a(x,\theta,\bar\theta)=\Sigma_a$ is complex linear,
    $\zeta^n_L(x,\theta,\bar\theta)=X_L^n$,  $\pi_{Ln}(x,\theta,\bar\theta)= Y_{Ln}$
    are left-semi-chiral and
    $\zeta_{R}^{n'}(x,\theta,\bar\theta)= X^{n'}_R$, $\pi^{Rn'}(x,\theta,\bar\theta)= Y_{Rn'}$ are right-semi-chiral
    specify a submanifold of ${\cal M} $ that  is  a holomorphic isotropic submanifold with respect to $\Omega$.
 
  As in the previous cases, enlarging the space of supermaps by introducing auxiliary fermionic superfields  $\Psi$
   should give a formulation in which the 
  submanifold (of the enlarged space) specified by the  superfield constraints   
    is Lagrangian. This  construction is straightforward locally. Globally,
    a  problem which may arise with the $\Psi$ fields is that of defining them 
     globally since we need to understand the global structures associated with our construction and this may require extra input 
      from the geometry, in particular a better understanding of the global issues related to symplectic realisations.

     There is another interesting case to consider consisting of the system (\ref{general-eqmot}) with no $(\phi, \Sigma)$ superfields. 
   If we write  the general field redefinitions  (\ref{superf-transf}) for this specific case then we have
\bea 
 &&\chi~\longrightarrow~f_5(\chi)~,\nn  \\
 && \Lambda ~\longrightarrow~ f_6(\chi) \Lambda + f_7 ( \chi, X_L) + f_8 ( \chi, \bar{X}_R)  ~,\nn \\
 && X_L ~\longrightarrow~ f_9(\chi, X_L,Y_L)~,~~~Y_L ~\longrightarrow~ \tilde{f}_9(\chi, X_L,Y_L)~ , \nn \\
&& \bar{X}_R~\longrightarrow~
{f}_{10}(\chi, \bar{X}_R, \bar{Y}_R)~,~~~ \bar{Y}_R~\longrightarrow~
{\tilde{f}}_{10}(\chi, \bar{X}_R, \bar{Y}_R)~.\nn
 \eea
  In these transformations the superfields $(\chi, \Lambda, X_L, Y_L, \bar{X}_R, \bar{Y}_R)$  do not mix  with superfields
   $(\bar{\chi}, \bar{\Lambda}, \bar{X}_L, \bar{Y}_L, X_R, Y_R)$.  Thus we can claim that the set $(\chi, \Lambda, X_L, Y_L, \bar{X}_R, \bar{Y}_R)$ 
    corresponds to holomorphic coordinates and  we again have a well-defined holomorphic symplectic structure 
    \be
     \Omega =  \int d^2x~ d^4 \theta ~  \Big  ( \delta \Phi_{a'}  \wedge \delta \Phi^{a'} +  \delta \Phi_{n}  \wedge \delta \Phi^{n} 
     + \delta \bar{\Phi}_{\bar{n}'}  \wedge \delta \bar{\Phi}^{\bar{n}'}  \Big )
    \ee
     and the discussions from the previous case go through similarly, after interchanging $n' \leftrightarrow {\bar n}'$. Thus the system (\ref{general-eqmot}) without $(\phi, \Sigma)$
      can again be interpreted as the intersection of real Lagrangian and holomorphic isotropic (or Lagrangian) submanifolds but with a modified 
       holomorphic symplectic structure.

         These two cases, either without $(\chi, \Lambda)$ or without $(\phi, \Sigma)$, have target spaces which are generalised K\"ahler manifolds 
   of symplectic type and the corresponding geometry has been studied in \cite{Bischoff:2018kzk}.  
     For any generalised K\"ahler manifold
   of symplectic type 
   a \lq doubled space' was constructed using a holomorphic Morita equivalence in 
     \cite{Bischoff:2018kzk} that is
     a holomorphic symplectic manifold. Here we have also constructed a \lq doubled'
        holomorphic symplectic manifold $({\cal Z}, \omega)$ and 
        we conjecture that the two constructions in fact agree.
          As far as we can see, the present discussion is consistent with the global considerations
    in \cite{Bischoff:2018kzk}.

\subsection{Chiral and twisted chiral case}   
  
  We now consider the case of generalised K\"ahler geometry for which the corresponding sigma model  is formulated only in terms of chiral and twisted chiral fields. 
  We again  investigate the global structure of the system  (\ref{general-eqmot})  without $X$ and $Y$ fields by 
 examining  the symmetries of these equations.  
 The chirality conditions allow the field redefinitions $\phi^a\rightarrow {\phi'}^a(\phi)$, $\chi^{a'}\rightarrow \chi'^{a'} (\chi)$ and the
  linear complex superfields $\Sigma_a$ and  twisted linear complex superfields $\Lambda_{a'}$  transform under these 
  coordinate transformations as
    \be
     \Sigma_a \rightarrow   \Sigma'_a=\frac{\partial \phi^b}{\partial{\phi'}^a}~  \Sigma_b ~,~~~~~
       \Lambda_{a'} \rightarrow   \Lambda'_{a'}=\frac{\partial \chi^{b'}}{\partial{\chi'}^{a'}}~  \Lambda_{b'} ~.
    \ee
    Note that this 
     is compatible with the superfield structure (see Appendix A). In addition to these symmetries we have the following glueing of $\Sigma$'s and $\Lambda$'s
     \bea
&&    ( \Sigma_a)_\beta  =
( \Sigma_a )_\alpha + \frac{\partial}{\partial \phi^a} \Big ( F_{\alpha\beta}^+ (\phi, \chi)  +   F_{\alpha\beta}^- (\phi, \bar{\chi})  \Big )~, \label{gen-shift-1-special} \\
  &&    ( \Lambda_{a'} )_\beta  =
(  \Lambda_{a'} )_\alpha 
+ \frac{\partial}{\partial \chi^{a'}} \Big ( F_{\alpha\beta}^+ (\phi, \chi )  + \bar{F}_{\alpha\beta}^-(\bar{\phi}, \chi ) \Big )~.  \label{gen-shift-2-speacial}
\eea
  The structure here is again a deformation ${\cal Z}$ of the cotangent bundle of the generalised K\"ahler manifold $M$, generalising that considered in the previous  section. 
   However there is an important difference as we do not have a holomorphic structure on ${\cal Z}$ since the above transformations mix barred and unbarred fields. 
  For   a patch $U_\alpha$ of $M$ (with $2d$   the real dimension of $M$),
we introduce  $d$ local complex coordinates  $Z^A_\alpha=(z^a_\alpha ,w^{a'}_\alpha)$ (corresponding to the lowest components of the superfields
$\varphi ^A= (\phi^{a},\chi^{a'})$). 
Then for the cotangent bundle
$T^*(U_\alpha)=U_\alpha\times \mathbb{R}^{2d}$  
we inroduce $d$ complex fibre coordinates
$(P_A)_\alpha=( (p _a))_\alpha ,
   (r_{a'} )_\alpha)$
(corresponding to the lowest components of the superfields
$\varphi _A=( \Sigma _a ,
   \Lambda_{a'} )$). 
 We then construct the deformed cotangent bundle ${\cal Z}$
by glueing together the patches $T^*(U_\alpha)$ with the following transition conditions in 
$T^*(U_\alpha\cap U_\beta)$:
\bea
&&    ( p_a)_\beta  =
( p_a )_\alpha + \frac{\partial}{\partial z^a} \Big ( F_{\alpha\beta}^+  +   F_{\alpha\beta}^-   \Big )~, \label{gen-shift-1-ex} \\
  &&    ( r_{a'} )_\beta  =
(  r_{a'} )_\alpha 
+ \frac{\partial}{\partial w^{a'}} \Big ( F_{\alpha\beta}^+    + \bar{F}_{\alpha\beta}^-  \Big )~,  \label{gen-shift-2-ex} 
    \eea
where the   functions $F^\pm$  have the following dependence on the  coordinates:
\be F_{\alpha\beta}^+  =F_{\alpha\beta}^+ (z^a  ,w^{a'})~,
\qquad
F_{\alpha\beta}^-   =F_{\alpha\beta}^- (z^a  ,\bar w^{{\bar a}'} )~.
\ee
 The properties of the glueing  in (\ref{GKP-ambig}) were studied in  \cite{Hull:2008vw}  where it was shown that  they are ultimately related to gerbes. One    consequence of 
  \cite{Hull:2008vw} is that the above glueing relations  (\ref{gen-shift-1-ex})-(\ref{gen-shift-2-ex}) satisfy the coycle conditions in triple overlaps and thus they are consistent glueing rules for a bundle (see Appendix B).

In each patch $T^*(U_\alpha)$ there is, as the notation suggests, a complex structure that acts as $+i$ on $(dZ^A, dP_A)$, together with
a holomorphic
symplectic structure
\be
\omega^{(2,0)} = dP_A\wedge dZ^A ~.
\ee
However,  the transition functions (\ref{gen-shift-1-ex})-(\ref{gen-shift-2-ex}) mix the holomoprhic coordinates $(Z^A,P_A)$ with the antiholomorphic ones $(\bar Z^{\bar A},\bar P_{\bar A})$
so the local complex structure and holomorphic
symplectic structure do not extend to holomorphic structures on ${\cal Z}$.
Thus there is no {\emph {manifest}}   complex structure on ${\cal Z}$ in 
       general, although of course the generalised K\"ahler manifold has two complex structures $I_\pm$.
       The bundle is then a real bundle with fibres $\mathbb{R}^{2d}$ and can't be viewed as a holomorphic bundle using the  complex structure on each patch.
However, there is a well-defined real symplectic form
\be
2~ {\rm Re}(\omega^{(2,0)}) = dP_A\wedge dZ^A + d\bar P_{\bar A}\wedge d\bar Z^{\bar A}~.
\ee
Explicit calculation shows that this symplectic form  is invariant under  the transformations in the transition functions (\ref{gen-shift-1-ex})-(\ref{gen-shift-2-ex}), 
so that they are symplectomorphisms.
Then ${\rm Re }(\omega) $ is globally well-defined on
the deformed cotangent bundle ${\cal Z}$. In this way we obtain an affine symplectic bundle ${\cal Z}$. 
 The transition functions here are closely related to gerbes, see  \cite{Hull:2008vw}.  
 Moreover, it is not clear how the space ${\cal Z}$ 
  encodes generalised K\"ahler geometry of $M$ (this is known for the previous examples). 
 
 Next we  define the  infinite dimensional space of  
   supermaps from $\mathbb{R}^{2|4}$ to the
  symplectic affine bundle constructed above
  \be
   {\cal M} = \{ \Phi :\mathbb{R}^{2|4}\longrightarrow {\cal Z} \}~,
  \ee
   which is equipped with the real symplectic form 
   \be
    2{\rm Re}( \Omega ) =   \int d^2x~ d^4 \theta ~  \Big (  \delta \Phi_a  \wedge \delta \Phi^a +  \delta \Phi_{a'}  \wedge \delta \Phi^{a'} +
    \delta \Bar{\Phi}_{\bar{a}}  \wedge \delta \bar{\Phi}^{\bar{a}} +  \delta \bar{\Phi}_{\bar{a}'}  \wedge \delta \bar{\Phi}^{\bar{a}'} \Big )~.
   \ee
   The condition
    \be
    \Phi_A = \frac{\partial K}{\partial \Phi^A}
   \ee
   defines a real Lagrangian submanifold with respect to  ${\rm Re}( \Omega )$.  The superfields constraints with their complex conjugate 
   define a real isotropic submanifold. 
    In analogy with K\"ahler case we can introduce introduce fermionic fields and define the corresponding generating functions so that the subspace defined by the superfield constraints is a real Lagrangian submanifold.

\subsection{Geometric Structure of General Case}
\label{GeomGen}

We now consider the target space of the of general sigma model with chiral, twisted chiral and semi-chiral superfields.
We introduce local coordinates $Z^A = (z^a,w^{a'},\zeta_L ^n,  \zeta_R^{n'})$ on the manifold $M$ corresponding to the lowest components of the superfields $\varphi^A = (\phi^a,\chi^{a'},X_L^{n}, X_R^{n'})$
together with dual coordinates $\pi_{Ln}, \pi_{Rn'}$, 
corresponding to the lowest components of the superfields $Y_{Ln}, Y_{Rn'}$,
that are defined  by
\be
  \pi_L=\frac { \partial K}{ \partial \zeta_L}~,
  \qquad
  \pi_R=\frac { \partial K}{ \partial  \zeta_R}~,
  \ee
where $K(Z,\bar Z)$ is the generalised K\"ahler potential.
Then $z,w,\zeta_L,\pi_L$ are $I_+$-holomorphic coordinates on $M$ and
$z,\bar w,\zeta_R,\pi_R$ are $I_-$-holomorphic coordinates on $M$.
On the overlap  $U_\alpha \cap U_\beta$ of two patches on $M$, (\ref{GKP-ambig}) gives the glueing conditions for the generalised K\"ahler potential  in the two patches to be
\be\label{GKP-ambig2}
    K_\alpha - K_\beta = F_{\alpha\beta}^+ (z,w, \zeta_L) + \bar{F}^+(\bar{z}, \bar{w}, \bar{\zeta}_L) + 
     F_{\alpha\beta}^- (z, \bar{w},  \zeta_R) + \bar{F}_{\alpha\beta}^-(\bar{z}, w, \bar{ \zeta}_R)~.
   \ee
Note that $F^+$ is $I_+$-holomorphic and $F^-$ is $I_-$-holomorphic.

This can be recast
in terms of  a product space of the kind considered in the previous subsections:
  \be
        {\cal N}= M_+\times M_- ~,
        \ee
        where $M_+=(M,I_+)$ and $M_-=(M,I_-)$ and $ {\cal N}$ has complex structure
       $  I=I_+\times I_-$.    
       Taking $z,w,\zeta_L,\pi_L$  as $I_+$-holomorphic coordinates on $M_+$ and
$z', w',\zeta_R,\pi_R$ as $I_-$-holomorphic coordinates on $M_-$, the manifold $M$ with coordinates $z^a,w^{a'},\zeta^n_{R}, 
 \zeta^{n'}_R$
 is obtained as the submanifold specified by
\be
z'=z~,  \qquad
w'=\bar{w}~ , \qquad
  \pi_L=\frac { \partial K}{ \partial \zeta_L}~,
  \qquad
  \pi_R=\frac { \partial K}{ \partial  \zeta_R}~.
  \label{doubeq}
  \ee
  Note the twist:
  $z'$ is identified with $z$ but $w'$ is identified with the {\emph {complex conjugate}} of $w$.

 The space $M_+$ is foliated 
 by leaves on which $z$ and $w$ are constant, with coordinates $\zeta_L,\pi_L$ on each leaf.
 Similarly, $M_-$ is foliated 
 by leaves on which $z'$ and $w'$ are constant, with coordinates $\zeta_R,\pi_R$ on each leaf.
 Introducing a relation $z\sim z'$, $w\sim \bar{w}'$, then taking the quotient 
     \be
      {\hat M}= {\cal N}/\sim
      \ee
      gives a space with coordinates $z,w,\zeta_L,\pi_L,\zeta_R,\pi_R$, which can be thought of as the original space $M$ with \lq doubled' leaves, which agrees with the space $\hat M$ from subsection \ref{semichichi} in the case in which there are no twisted chiral fields and hence no coordinates $w$. Much of the discussion from subsection \ref{semichichi} extends to this case.

We now  construct a bundle ${\cal Z}$ over $\hat M$ with fibre coordinates $p_a, r_{a'}$ and
        glueing conditions        
         \bea\label{pr-glue-stype}
      p'_a&=&p_a + \frac {\partial} {\partial z^a}
        \left[ F^+(z,w,\zeta_L)+ F^-(z,\bar w,\zeta_R) \right]~,
        \\
        r'_{a'}&=&r_{a'} + \frac {\partial} {\partial w^{a'}}
        \left[ F^+(z,w,\zeta_L)+ 
        \bar F^-(\bar z,w,\zeta_R) \right]~.
        \eea
             It is important that these satisfy the cocycle condition in triple overlaps (see Appendix B) so that this is indeed a fibre bundle.
      With these glueing rules, the section of      ${\cal Z}$  defined by
      \be
      p_a = \frac {\partial} {\partial z^a}K, 
      \qquad
      r_{a'} = \frac {\partial} {\partial w^{a'}} K
      \ee 
      is a global section defining a submanifold.
      
      This is then the geometric setting for the equations (\ref{general-eqmot}).
         Globally we are able to define only a real symplectic structure  and there is no natural complex structure.
    Thus at best we can interpret the equations (\ref{general-eqmot}) as the defining intersection of two real Lagrangian submanifolds in the space of unconstrained 
     superfields. In the next section we present an alternative formulation which does have a holomorphic structure.


\section{Extended geometry}\label{s:extended}

In the  previous sections we  discussed the symplectic interpretation of the equations (\ref{general-eqmot}). In certain special
cases we have a superfield phase space equipped with a complex structure and a holomorphic symplectic form. However, this is 
 not the case in the general situation. We  seek here  an extension of   the target space
   that  carries a holomorphic symplectic structure. 
   We find a surprising  reformulation that has an  interesting geometry and
   is dictated by the superfield structure.

\subsection{Dual superfield presentation}

The (twisted) complex linear superfields can be re-expressed in terms of semi-chiral superfields.
 The complex linear constraints    
               \be
 \bar{\mathbb{D}}_+  \bar{\mathbb{D}}_- \Sigma_a=0~,
 \qquad
 \bar{\mathbb{D}}_+ \mathbb{D}_- \Lambda_{a'}=0
 \ee
are solved by
\be
\label{solev}
\Sigma_a= \bar{\mathbb{D}}_+ \Psi^+_a+ \bar{\mathbb{D}}_- \Psi^+_a~,
~~~~~
\Lambda_{a'}=\bar{\mathbb{D}}_+ \Upsilon _{a'}^++ \mathbb{D}_-\Upsilon _{a'}^-
\ee
for some unconstrained 
spinor superfields $\Psi^\pm_a,\Upsilon _{a'}^\pm $.
Then  (twisted) complex linear superfields  can   be rewritten as
\be
\label{solev}
\Sigma_a= U_{La}   
+ U_{R a} ~,
\qquad
\Lambda_{a'}=V_{L a'}
+ \bar V_{R a'}  ~,
\ee
where
\be U_{La}  = \bar{\mathbb{D}}_+ \Psi^+_a, \qquad U_{R a} = \bar{\mathbb{D}}_- \Psi^+_a,
\ee
\be
V_{L a'}=
\bar{\mathbb{D}}_+ \Upsilon _{a'}^+, \qquad
 \bar V_{R a'} =\mathbb{D}_-\Upsilon _{a'}^-
\ee
are semi-chiral superfields.
As we shall see, formulating the theory in terms of the semi-chirals $U_L,U_R,V_L,V_R$
instead of the linear superfields $\Sigma, \Lambda$ is very helpful. 
 Using this we can rewrite the equations  (\ref{general-eqmot}) as follows
 
\fbox{
 \addtolength{\linewidth}{-2\fboxsep}%
 \addtolength{\linewidth}{-2\fboxrule}%
 \begin{minipage}{\linewidth}
\bea\label{extended-general-eqmot}
 && U_{La} + U_{Ra}= \frac{\partial K}{\partial \phi^a}~,~~~~V_{La'} + \bar{V}_{Ra'} =  \frac{\partial K}{\partial \chi^{a'}}~,~~~~
    Y_{Ln}=  \frac{\partial K}{\partial X^n_L}~,~~~~Y_{Rn'} =  \frac{\partial K}{\partial X^{n'}_R} \nonumber \\ 
&& \bar{\mathbb{D}}_+ U_{La}=0 ~,~~~\bar{\mathbb{D}}_- U_{Ra}=0 ~,~~~
 \bar{\mathbb{D}}_+ V_{La'}=0 ~,~~~\mathbb{D}_- \bar{V}_{Ra'}=0 ~, 
 ~~~\bar{\mathbb{D}}_+ Y_{Ln}=0~,~~~~\bar{\mathbb{D}}_- Y_{Rn'}=0 \nonumber \\
 &&    \bar{\mathbb{D}}_{\pm} \phi^a=0~,~~~\bar{\mathbb{D}}_+ \chi^{a'} = \mathbb{D}_- \chi^{a'} =0~,~~~\bar{\mathbb{D}}_+ X^n_L=0~,~~~~\bar{\mathbb{D}}_- X^{n'}_R=0 
\eea
 \end{minipage}
}

Let us discuss the symmetries of these equations.  From the glueing conditions  
(\ref{gen-shift-1}),(\ref{gen-shift-2}) we get the following glueing
  in terms of the new variables   
\bea
&&    ( U_{La})_\beta  =
( U_{La} )_\alpha + \frac{\partial}{\partial \phi^a} F_{\alpha\beta}^+ (\phi, \chi, X_L)     ~, \label{gen-shift-5} 
\\
 &&    ( U_{Ra})_\beta  =
(U_{Ra} )_\alpha + \frac{\partial}{\partial \phi^a}  F_{\alpha\beta}^- (\phi, \bar{\chi}, X_R)   ~, \label{gen-shift-6} 
\\
  &&    ( V_{L a'} )_\beta  =
( V_{L a'} )_\alpha 
+ \frac{\partial}{\partial \chi^{a'}}  F_{\alpha\beta}^+ (\phi, \chi, X_L)    ~,  \label{gen-shift-7} 
\\
  &&    ( \bar V_{R a'} )_\beta  =
(  \bar V_{R a'} )_\alpha 
+ \frac{\partial}{\partial \chi^{a'}} \bar{F}_{\alpha\beta}^-(\bar{\phi}, \chi, \bar{X}_R)   ~,  \label{gen-shift-8} 
    \eea
whereas the transformations (\ref{superf-transf-double}) for $\Sigma$ and $\Lambda$ become
       \bea
&& U_L~\longrightarrow~f_2(\phi)U_L + f_3 (\phi, \chi, X_L, Y_L)  ~,\nn \\
 && U_R~\longrightarrow~f_2(\phi)U_R  +  f_4(\phi, \bar{\chi}, X_R, Y_R) ~,\nn \\
 && V_L ~\longrightarrow~ f_6(\chi) V_L+ f_7 (\phi, \chi, X_L, Y_L)  ~,\nn \\
 && \bar V_{R } ~\longrightarrow~ f_6(\chi) \bar V_{R }  + f_8 (\bar{\phi}, \chi, \bar{X}_R, \bar{Y}_R)  ~.\label{gentransy}  
 \eea
 The glueing conditions and transformations for 
 $U_L,V_L$ are $I_+$-holomorphic and those for $U_R,V_R$ are $I_-$-holomorphic.
  This would be consistent with $(U_L,V_L)$
  being fibre coordinates for some holomorphic  bundle over $M_+$ and 
$(U_R,V_R)$ being fibre coordinates for some  holomorphic  bundle over $M_-$.
Moreover, the transformations (\ref{gentransy}) are not the most general ones consistent with the chirality constraints and their special form is indicative of a role as fibre coordinates. 
However, this is not the correct picture.
The crucial point is that (\ref{gen-shift-5})-(\ref{gen-shift-8}) do not satisfy the cocycle conditions on triple intersections
(see Appendix B) so that there is no interpretation of $(U_L,V_L)$ or $(U_R,V_R)$  as extra coordinates for an enlarged manifold. Instead, adding the extra variables gives rise to a more general structure. We discuss this   further below.

\subsection{Geometric Structure}

In subsection \ref{GeomGen}, we considered the manifold ${\cal N}= M_+\times M_- $ with complex structure
       $  I=I_+\times I_-$.    The manifold $M$ was recovered as the submanifold specified by (\ref{doubeq}).
The space ${\hat M}$ with coordinates $z,w,\zeta_L,\pi_L,\zeta_R,\pi_R$ (obtained from the  space $M$ by doubling the leaves) was constructed as the quotient ${\hat M}= {\cal N}/\sim$ of $\hat M$ by the relation $z\sim z'$, $w\sim \bar{w}'$.
We then  constructed a bundle ${\cal Z}$ over $\hat M$ with fibre coordinates $p_a, 
r_{a'}$ and
        glueing conditions        
    (\ref{pr-glue-stype}), but this did not have a natural complex structure.

We now introduce   variables
$p_{La} ,p_{R a}, r_{L a'},r_{R a'}
$
 corresponding to the lowest components of the superfields $U_{La} ,U_{R a}, V_{L a'},V_{R a'}
$. 
The relations
(\ref{solev})  give
 \be
 \label{pplr}
    p=p_L+p_R~,
    \qquad r=r_L+\bar r_R~.
    \ee
It would be natural to attempt to generalise the previous constructions by interpreting $p_L,r_L$ as fibre coordinates for some  bundle $E_+$ over $M_+$ and $p_R,r_R$ as fibre coordinates for some  bundle $E_-$ over $M_-$ so that $E_+\times E_-$ has a natural complex structure and the bundle 
${\cal Z}$ can be constructed from it by taking the quotient by $\sim$ and using (\ref{pplr}). We shall develop this picture below, finding a natural holomorphic setting for our equations. 
However, as we shall see, the $E_\pm$ that arise in this way are {\emph {not}} fibre bundles and are instead generalised spaces.

For $M_+$ we use the $I_+$-holomorphic coordinates $z,w,\zeta_L,\pi_L$
 and for each patch $U_\alpha$ we introduce additional variables $p_L,r_L$ parameterising local fibres of a bundle $V_\alpha$ over $U_\alpha$.
 Over intersections  $U_\alpha\cap U_\beta$,   we have, from 
(\ref{gen-shift-5}),(\ref{gen-shift-7}), the transition functions
 \bea
&&    ( p_{La})_\beta  =
( p_{La} )_\alpha + \frac{\partial}{\partial z^a} F_{\alpha\beta}^+ (z, w, \zeta_L)    ~, \label{gen-shift-9} 
\\
&& ( r_{La'})_\beta  =
( r_{La'} )_\alpha + \frac{\partial}{\partial w^{a'}} F_{\alpha\beta}^+ (z, w, \zeta_L)    ~, \label{gen-shift-10} \\
&& (\pi_{Ln})_\beta =  (\pi_{Ln})_\alpha + \frac{\partial}{\partial \zeta_{L}^{n}} F_{\alpha\beta}^+ (z, w, \zeta_L)~
   \eea
The $V_\alpha$ are then glued using the  $I_+$-holomorphic glueing relations (\ref{gen-shift-9}),(\ref{gen-shift-10}) to construct some \lq generalised space' $E_+$. 
Locally,    $p_{La}, r_{L a'}
$  are fibre coordinates for a bundle $V_\alpha$ over a patch $U_\alpha$.
However, globally this does not extend to a bundle over $M_+$, or indeed any picture in which these variables are coordinates of some manifold.
This is because the transition functions for $p_{La} ,r_{L a'}
$ do not satisfy the cocycle condition in triple overlaps (see Appendix B) and so these variables cannot be interpreted as coordinates on any bundle over $M_+$.
Then globally, $E_+$ is not a bundle and not a manifold.

 We proceed formally  and note that we  can introduce a complex structure and 
       symplectic structure 
      on each $V_\alpha$. These are 
       invariant under the glueing relations and so extend to structures on $E_+$. 
  Then  $E_+$ has a complex structure $I_+$ for which $z,w,p_L,r_L,\zeta_L,\pi_L$
are holomorphic variables, and an $I_+$-holomorphic symplectic structure  
  \be
            \omega^{(2,0)}_+ =  d\pi_{Ln} \wedge d\zeta^n _L
            +d  p_{La}    \wedge  d z^a
            +d r_{La'}  \wedge     dw^{a'}~,
           \ee
      which is invariant under the  glueing relations.

     Similarly, we take $p_R,r_R$ to be additional variables for a generalised space  $E_-$ over $M_-$ 
     From 
(\ref{gen-shift-6}),(\ref{gen-shift-8}) we have the  transition functions  
\bea
&&    ( p_{Ra})_\beta  =
(p_{Ra} )_\alpha + \frac{\partial}{\partial {z'}^a}  F_{\alpha\beta}^- (z', w', \zeta_R)  ~, \label{gen-shift-11} \\
  &&    (  r_{R a'} )_\beta  =
(   r_{R a'} )_\alpha  + \frac{\partial}{\partial w'^{a'}} F_{\alpha\beta}^-(z', w', \zeta_R) ~,  \label{gen-shift-12} \\
&&   (\pi_{Rn'})_\beta =  (\pi_{Rn'})_\alpha + \frac{\partial}{\partial \zeta_{R}^{n'} } F_{\alpha\beta}^- (z', w', \zeta_R)~.
    \eea
     This has simliar properties to $E_+$, and for the same reasons $E_-$  does not define a manifold.
However, $E_-$ has a complex structure $I_-$, for which $z', w',p_R,r_R, \zeta_R, \pi_R$
are holomorphic, and it has a $I_-$-holomorphic symplectic structure
  \be
            \omega^{(2,0)}_- =  
            d \pi_{Rn'} \wedge d \zeta^{n'}_R
            +d  p_{Ra}    \wedge  d {z'}^a
            +d r_{Ra'}  \wedge     d w'^{a'}
            ~,
           \ee
    which is invariant under the  glueing.

To understand the geometry of the construction, recall that for the general (2,2) sigma models the 3-form $H$ represents an integral cohomology class  in the quantum theory
and is the curvature for a gerbe on $M$. The interplay between the  gerbe strucure  and the complex structures was explored in  \cite{Hull:2008vw} where it was found that a rich system
 of holomorphic gerbes underpins the theory. In particular, the transition functions $F^+_{\alpha \beta}$ and $F^-_{\alpha \beta}$ defined in  intersections do not satisfy the cocycle condition that 
 $\delta F^\pm\equiv F_{\alpha\beta}^\pm +  F_{\beta\gamma}^\pm  +  F_{\gamma\alpha}^\pm $ vanishes (or vanishes modulo $2\pi$ when appropriately normalised) but instead $\delta F^+$ and $\delta F^-$ each provides a non-trivial map from  triple intersections $U_\alpha \cap U_\beta \cap U_\gamma$ to $U(1)$ that defines a gerbe structure. 
 Moreover, as $F^+$ is $I_+$-holomorphic and $F^-$ is $I_-$-holomorphic, these are holomorphic gerbes.
 See Appendix B for more details.
In our construction, 1-forms $p_{La} , r_{L a'} $
 are introduced on each patch
 of $M_+$ and
 glued together using the gerbe transition functions through (\ref{gen-shift-9}),(\ref{gen-shift-10}),
 while $p_{R a},r_{R a'}$
   are introduced on each patch
 of $M_-$
  and glued   using  (\ref{gen-shift-11}),(\ref{gen-shift-12}). In fact, they are glued in the same way as certain prepotentials for holmorphic gerbe connections (see Appendix B). This can be viewed as a generalisation of the Donaldson construction, where (\ref{KG-momglform}) implies that $p$ has the glueing relations of a holomorphic connection.
 
 The advantage of this construction is its manifest holomorphic structure and its holomorphic symplectic structure.
 We introduce the space
      \be 
      {\cal X}=E_+\times E_-~,
      \ee
     which has a complex structure $I=I_+\times I_-$ and a $I$-holomorphic symplectic structure
      $\omega^{(2,0)}=\omega^{(2,0)}_+ + \omega^{(2,0)}_-$.
      The $I$-holomorphic coordinates $\hat Z^A$ and conjugate variables $\hat P_A$ are
      \be
        \hat Z^A=(z,w,z',w',\zeta_L,\zeta_R)~, \qquad \hat P_A=(p_L,r_L,p_R,r_R,\pi_L,\pi_R)~.
        \label {hatzpis}
        \ee
        Then the $I$-holomorphic symplectic structure can  be written as
        \be \label {sigmis}
        \omega^{(2,0)}=
        d\hat P_A\wedge d \hat{Z}^A~.
        \ee
      This   space   is an extended arena for our equations. In particular, our equations lead to the subspace of this given by
    (\ref{doubeq}) together with
      \be
      p_L+p_R =\frac { \partial K}{ \partial z}~
      ,\qquad
      r_L+\bar r_R=\frac { \partial K}{ \partial w}~.
            \ee

    Recall that  the manifold $M$   is obtained as the submanifold of $M_+\times M_-$ specified by (\ref{doubeq}).
Then from $ {\cal X}$ we obtain a subspace in which over each point in this submanifold $M$ we have  fibre coordinates
$p_L,r_L , p_R,r_R$. We now define
       \be
    p=p_L+p_R~,
    \qquad r=r_L+\bar r_R~.
    \ee
and from 
       (\ref{gen-shift-9}),(\ref{gen-shift-10}),(\ref{gen-shift-11}),(\ref{gen-shift-12})  the transition functions for   $p,r$   are
    \bea
&&    ( p_a)_\beta  =
( p_a )_\alpha + \frac{\partial}{\partial z^a} \Big ( F_{\alpha\beta}^+  +   F_{\alpha\beta}^-   \Big )~, \label{gen-shift-1-ex-ex} \\
  &&    ( r_{a'} )_\beta  =
(  r_{a'} )_\alpha 
+ \frac{\partial}{\partial w^{a'}} \Big ( F_{\alpha\beta}^+    + \bar{F}_{\alpha\beta}^-  \Big )~,  \label{gen-shift-2-ex-ex} 
    \eea
where the   functions $F^\pm$  have the following dependence on the  coordinates:
\be F_{\alpha\beta}^+  =F_{\alpha\beta}^+ (z  ,w,\zeta_L)~,
\qquad
F_{\alpha\beta}^-   =F_{\alpha\beta}^- (z   ,\bar w  ,\zeta_R)~.
\ee
These  glueing relations do satisfy the cocycle condition in triple overlaps (see Appendix B) and so define a bundle over $M$, with coordinates $z,w,\zeta_L, \zeta_R$ on $M$ and fibre coordinates $p,r$.
As the functions $F^\pm$ are independent of $\pi_L,\pi_R$, this bundle in fact extends to a bundle   
      $\hat {\cal Z}$ over $\hat M$ with fibre coordinates $p,r$ 
      and coordinates $z,w,\zeta_L,\pi_L,\zeta_R,\pi_R$ on $\hat M$.

     The space $\hat {\cal Z}$  inherits a real symplectic structure from $\omega^{(2,0)}$ given by
      \be
        \omega=
        dp_a\wedge dz^a + dr_{a'}\wedge d\bar w^{a'}+d\pi_{Ln}\wedge d \zeta _L^n
        +\pi_{Rn'}
        \wedge d \zeta^{n'}_R + {\rm complex~conjugate}
        \ee
        With the notation
        \be
        Z^A=(z,w,\zeta_L,\zeta_R), \qquad P_A=(p,r,\pi_L,\pi_R)
        \label {zpis}
        \ee
        this symplectic structure can be written as
        \be \label {omis}
        \omega =
        dP_A\wedge d Z^A + d\bar{P}^{\bar{A}} \wedge d\bar{Z}^{\bar{A}}~. 
        \ee
        However, ${\cal Z}$ does not inherit a complex structure in general. 
        In the special case in which there are no coordinates $w,r$ this does inherit a complex structure from $I$ for which 
$z,p,\zeta_L,\pi_L,\zeta_R,\pi_R$ are holomorphic coordinates; this is the case discussed in section \ref{semichichi}, and similarly for the case with no coordinates $z,p$.
In general, the symplectic structure $\omega$ of  $  {\cal Z}$ is not holomorphic but descends from a holomorphic symplectic structure on $  {\cal N}$.

The field equations specify the subspace of $  {\cal Z}$ 
\be
p_a=\frac { \partial K}{ \partial z^a}~,
  \qquad
  r_{a'}=\frac { \partial K}{ \partial w^{a'}}~
, \qquad
  \pi_L=\frac { \partial K}{ \partial \zeta_L}~,
  \qquad
  \pi_R=\frac { \partial K}{ \partial  \zeta_R}~,
  \label{doubeq2}
  \ee
which is Lagrangian with respect to $\omega$.

\subsection{An Extended Formulation}\label{ss:fields-4}

   We now consider supermaps to the extended space ${\cal X}$
   \be
   {\cal Q} = \{ \Phi :\mathbb{R}^{2|4}\longrightarrow {\cal X} \}~,
  \ee
  in order to interpret the superfield constraints as defining
  an isotropic  submanifold.
  The coordinates on ${\cal X}$ are
  $\hat Z^A,\hat P_A$ given by
  (\ref{hatzpis}) with holomorphic symplectic structure $ \omega^{(2,0)}=
        d\hat P_A\wedge d \hat{Z}^A$.
  The
  supermaps are
  unconstrained superfields $\hat\Phi^A=\hat Z^A(x,\theta,\bar
  \theta)$, $\hat\Phi_A=\hat P_A(x,\theta,\bar
  \theta)$ and this space has a holomorphic symplectic structure derived from (\ref{sigmis}), which is
  \be
 \hat \Omega =  \int d^2x~ d^4 \theta ~   \delta \hat\Phi_A  \wedge \delta \hat \Phi^A~. 
  \ee
  
  We now wish to formulate the superfield constraints as defining a submanifold of
  ${\cal Q} $.
  For $\zeta_L,\zeta_R,\pi_L,\pi_R,p_L,r_L,p_R,r_R$ we choose the constraints that correspond to  identifying these supermaps with
  $X_L,X_R,Y_L,Y_R,U_L,V_L,U_R,V_R$ respectively.
  For $z,z',w,w'$ we want constraints such that, on setting
  $z=z',w=\bar w'$, the supermap  $z=z'$ corresponds to the chiral superfield $\phi$ and 
  the supermap  $w=\bar w'$ corresponds to the twisted chiral superfield $\chi$.
  One way of doing this is to take $z,z'$ to be chiral and $w,w'$ to be twisted chiral.
  However, there is another possibility involving semi-chirals instead.
  We choose constraints that identify $z,z',w,w'$ with semi-chiral fields $Z,W$:
  \be
  z\sim Z_L~, \qquad z'\sim Z_R~, \qquad w\sim W_L~, 
  \qquad
  w'\sim  W_R~.
  \ee
  Then
  $z=z'$ corresponds to
  \be
  Z_L=Z_R
  \ee
  and the constraints on $Z_L,Z_R$ imply that $\phi\equiv Z_L=Z_R$ is a chiral superfield.
  Similarly, $w=\bar{w}'$ corresponds to
   \be
  W_L=\bar W_R
  \ee  
  and the constraints on $W_L,W_R$ imply that $\chi\equiv W_L=\bar W_R
$ is a twisted chiral superfield.

This gives a formalism in which \emph{all} superfields are semi-chiral.
The superfields are
\be
X_L,X_R,Y_L,Y_R,U_L,V_L,U_R,V_R,Z_L,Z_R,W_L, W_R
\ee
and the constraints
are

\fbox{
 \addtolength{\linewidth}{+0.9\fboxsep}%
 \addtolength{\linewidth}{-1\fboxrule}%
 \begin{minipage}{\linewidth}
\bea\label{dualy-eqmot-new}
  U_{La} = \frac{\partial \hat{K}}{\partial Z_L^a}~,~~Z_R^a =  \frac{\partial \hat{K}}{\partial U_{Ra}}~,~~
   V_{Ra'} = \frac{\partial \hat{K}}{\partial W_R^{a'}}~,~~W_L^{a'} =  \frac{\partial \hat{K}}{\partial V_{La'}}~,~~
    Y_{Ln}=  \frac{\partial \hat{K}}{\partial X^n_L}~,~~Y_{Rn'} =  \frac{\partial \hat{K}}{\partial X^{n'}_R} \nonumber \\ 
     \bar{\mathbb{D}}_+ U_{La} =0
     ~,~~~
      \bar{\mathbb{D}}_- U_{Ra} =0
     ~,~~~
     \bar{\mathbb{D}}_+ V_{L a'}=0~,~~~
      \bar{\mathbb{D}}_- V_{R a'}=0~,~~~
     \bar{\mathbb{D}}_+ Y_{Ln}=0~,~~~~\bar{\mathbb{D}}_- Y_{Rn'}=0 
     \nonumber \\
     \bar{\mathbb{D}}_+ X^n_L=0~,~~~~\bar{\mathbb{D}}_- X^{n'}_R=0 
   ~,~~~~   \bar{\mathbb{D}}_+ Z^a_L=0~,~~~~\bar{\mathbb{D}}_- Z^{a}_R=0~,~~~~
        \bar{\mathbb{D}}_+ W^{a'}_L=0~,~~~~\bar{\mathbb{D}}_- W^{a'}_R=0
        \nonumber\\
\eea
 \end{minipage}
}
~\\
 where we have defined 
\be
\hat{K} =   K(Z_L, \bar{Z}_L, \bar{W}_R, W_R, X_L, \bar{X}_L, X_R, \bar{X}_R) - 
U_{Ra} Z^a_L - \bar{U}_{R\bar{a}} \bar{Z}^{\bar{a}}_L - V_{La'} \bar{W}^{a'}_R - \bar{V}_{L\bar{a}'} W_R^{\bar{a}'}
   \ee
in terms of the generalised K\"ahler potential $K(\phi, \bar{\phi}, \chi, \bar{\chi}, X_L, \bar{X}_L, X_R, \bar{X}_R)$.
 One can easily check that the above equations are equivalent to the equations (\ref{extended-general-eqmot}).
  The two last lines in (\ref{dualy-eqmot-new}) are the semi-chirality constraints and  can be interpreted as defining a 
   holomorphic isotropic submanifold of  ${\cal Q} $ with respect to $\hat \Omega $. The first line in  (\ref{dualy-eqmot-new})
    defines a real Lagrangian submanifold of  ${\cal Q} $ with respect to ${\rm Re}(\hat \Omega)$. The interesting new feature of this 
     extended construction is that the chirality constraints for chiral and twisted chiral superfields appear as emergent constraints 
      from the intersection of a Lagrangian submanifold with an isotropic one. 

 The   ambiguities (\ref{GKP-ambig}) in the definition of $K$ are now realised as 
  holomorphic diffeomorphisms 
   \bea
   && U_L~\rightarrow~U_L + \partial_{Z_L} F^+ (Z_L, W_L, X_L) ~,\nonumber\\
   &&  V_L ~\rightarrow~V_L + \partial_{W_L} F^+ (Z_L, W_L, X_L)~,\nonumber \\
    &&  Y_L~\rightarrow~Y_R + \partial_{X_L} F^+ (Z_L, W_L, X_L) ~,\\
    &&  V_R ~\rightarrow~V_R + \partial_{W_R} F^- (Z_R, W_R, X_R) ~,\nonumber\\
  &&  U_R~\rightarrow~U_R + \partial_{Z_R} F^- (Z_R,  W_R, X_R) ~, \nonumber\\
     && Y_R ~\rightarrow~Y_R + \partial_{X_R} F^- (Z_R, W_R, X_R)~, \nonumber
  \eea 
    which obviously respect the corresponding chirality constraints.

   \section{Doubly extended  space and superfields}\label{s:double}
   
   In the previous section we constructed an extension of the original superfield phase space  that 
      is equipped with a holomorphic symplectic form. The equations of motion  and constraints are equivalent to the intersection of a real 
     Lagrangian submanifold with a holomorphic isotropic submanifold. 
    In our   construction of ${\cal X}$, the set of coordinates $z,w$ were \lq quadrupled' to the set $z,z',w,w,p_L,p_R,r_L,r_R$ while the leaf  coordinates $\zeta_L,\zeta_R$ were doubled to $\zeta_L,\zeta_R,\pi_L,\pi_R$.
      In this section, we present an alternative formulation in which all coordinates are quadrupled.
          Here we present the superfield formulation and proceed formally.  
          Most likely the present construction is related to the double groupoid picture outlined in   \cite{Marco-paper} and  \cite{jiang}.

  \subsection{Model with  semi-chiral superfields only}\label{ss:semis-4}
  
  Let us start with the discussion of the case in which there are only semi-chiral fields. We follow the discussion and notation from   subsection 
  \ref{ss:only-semis}.  We  double the construction given   there and define a complex manifold 
  \be
  {\cal Z} \times {\cal Z} = M_+ \times M_- \times M_+ \times M_-
   \ee  
    with the holomorphic symplectic structure  
       \be
            \omega^{(2,0)} = dP_A \wedge dZ^A + dP^*_A \wedge dZ^{*A}  = 
            d\pi_{Ln} \wedge d\zeta^n_L+d\pi_{Rn'} \wedge d\zeta^{n'}_R +  d\pi^*_{Ln} \wedge d\zeta^{*n}_L+d\pi^*_{Rn'} \wedge d\zeta^{*n'}_R~,
           \ee
            where the coordinates with a star are the coordinates for the second   ${\cal Z}$. Doubling the coordinates also doubles the corresponding 
             superfields. We  use the same notation,   adding a star for  the superfields taking values in  the second   ${\cal Z}$. 
      For a given generalized K\"ahler potential $K(X_L, \bar{X}_L, X_R, \bar{X}_R)$ we define the following real generating function
      \be
     \hat{K}=  K(X_L, \bar{X}_L, X_R, \bar{X}_R) - X_L^{n} Y^*_{Ln} - \bar{X}_L^{\bar{n}} \bar{Y}^*_{L\bar{n}}  - X_R^{n'} Y^*_{Rn'}
        - \bar{X}_R^{\bar{n'}} \bar{Y}^*_{R\bar{n'}}~,
      \ee
       which depends  on exactly  half of the fields.  We   observe   that $\hat{K}$ defines the  submanifold
        specified by the following 
        equations
       \bea\label{semi-extended}
       && Y_{Ln} =  \frac{\partial \hat{K}}{\partial X_L^n}  =\frac{\partial K}{\partial X^n_L} - Y_{Ln}^* ~,\nonumber\\
       && X_L^{*n} =   \frac{\partial \hat{K}}{\partial Y^*_{Ln}}   = X^n_L~,\\
       && Y_{Rn'} =  \frac{\partial \hat{K}}{\partial X^{n'}_{R}}   = \frac{\partial K}{\partial X^{n'}_R} - Y_{Rn'}^*~, \nonumber \\
       && X_R^{*n'} =  \frac{\partial \hat{K}}{\partial Y_{Rn'}}=   X^{n'}_R~. \nonumber
\eea
If we make the shifts $Y_L \to  Y_L + Y_L^*$  and $Y_R \to  Y_R + Y_R^*$ we recover  
 the equations (\ref{semi-eqmot}). 
  
  The space of unconstrained  superfields
  \be
   \{ \Phi :\mathbb{R}^{2|4}\longrightarrow {\cal Z} \times {\cal Z} \}~,
  \ee
  is equipped with the holomorphic symplectic form
   \be
 \Omega =  \int d^2x~ d^4 \theta ~   \Big ( \delta \Phi_A  \wedge \delta  \Phi^A +  \delta \Phi^*_A  \wedge \delta  \Phi^{*A} \Big )~.
   \ee
    The semi-chiral constraints on the superfields correspond to a holomorphic isotropic submanifold with respect to $\Omega$. 
     On the other hand, the equations (\ref{semi-extended}) correspond to a  real Lagrangian submanifold with respect to ${\rm Re}(\Omega)$. 
      Thus we conclude that the original equations of motion and constraints can be interpreted as the intersection of these two   
       submanifolds.

\subsection{General case}

Formally,   the general case can be viewed as a combination of the constructions from subsections \ref{ss:fields-4}
 and  \ref{ss:semis-4}.  
For a   generalised K\"ahler potential  $K(\phi, \bar{\phi}, \chi, \bar{\chi}, X_L, \bar{X}_L, X_R, \bar{X}_R)$ we define 
\bea
\hat{K} =   K(Z_L, \bar{Z}_L, \bar{W}_R, W_R, X_L, \bar{X}_L, X_R, \bar{X}_R) - 
U_{Ra} Z^a_L - \bar{U}_{R\bar{a}} \bar{Z}^{\bar{a}}_L - V_{La'} \bar{W}^{a'}_R - \bar{V}_{L\bar{a}'} W_R^{\bar{a}'} \nonumber \\
- X_L^{n} Y^*_{Ln}  - \bar{X}_L^{\bar{n}} \bar{Y}^*_{L\bar{n}}  - X_R^{n'} Y^*_{Rn'}
        - \bar{X}_R^{\bar{n'}} \bar{Y}^*_{R\bar{n'}}~.
   \eea
    Then the equations of motion can be written as follows 
 
 \fbox{
 \addtolength{\linewidth}{-2\fboxsep}%
 \addtolength{\linewidth}{-2\fboxrule}%
 \begin{minipage}{\linewidth}
\bea\label{dualy-eqmot-semi}
  U_{La} = \frac{\partial \hat{K}}{\partial Z_L^a}~,~~~Z_R^a =  \frac{\partial \hat{K}}{\partial U_{Ra}}~,~~~
   V_{Ra'} = \frac{\partial \hat{K}}{\partial W_R^{a'}}~,~~~W_L^{a'} =  \frac{\partial \hat{K}}{\partial V_{La'}}~, \nonumber \\
 X_L^{*n} =   \frac{\partial \hat{K}}{\partial Y^*_{Ln}} ~,~~~
 X_R^{*n'} =  \frac{\partial \hat{K}}{\partial Y_{Rn'}}~,~~~
    Y_{Ln}=  \frac{\partial \hat{K}}{\partial X^n_L}~,~~~Y_{Rn'} =  \frac{\partial \hat{K}}{\partial X^{n'}_R}~, \nonumber \\ 
      \eea
 \end{minipage}
}
~\\
  provided that we impose the semi-chiral conditions on all fields. All local geometry and holomorphic symplectic structures follow
   from those in subsections \ref{ss:fields-4} and  \ref{ss:semis-4}. In this bigger space we can interpret the equations of motion as the intersection of 
    a holomorphic isotropic submanifold (imposing the semi-chiral conditions on the superfields) and the real Lagrangian submanifold defined by $\hat{K}$. As 
     previously discussed, the ambiguities in $K$  are now realized as diffeomorphisms in this bigger space.

\section{Summary and outlook}\label{s:summary}

 We have   taken a fresh look at   generalised K\"ahler geometry motivated by the  $N=(2,2)$ superfield formulation of the corresponding sigma model. 
 It has long been known that the $N=(2,2)$ supersymmetric sigma  model with K\" ahler target space has a dual formulation in terms of complex linear superfields and here we have presented the natural extension of this to the general $N=(2,2)$ supersymmetric sigma model with generalised K\" ahler target space, giving a new  formulation that is dual to the usual one in terms of chiral, twisted chiral and semichiral supermultiplets.
 The duality interchanges superfield equations of motion with the superconstraints on the superfields.
Instead of the usual formulation in terms of an action (together with a dual action),  we have developed a novel reformulation with a doubled 
target space that focuses on the superfield equations of motion and makes the duality manifest.
Constraining to a Lagrangian submanifold of the doubled  space implies the field equations. This might be viewed as a kind of superspace Hamiltonian formalism, with different dual formulations arising from different choices of polarisation.  
Whereas much earlier work has focused on the local geometry, here we pay particular attention to global issues and find an interesting generalisation of a construction due to Donaldson. 

Somewhat surprisingly, alternative   superfield formulations
 leads us to quadruple 
 some or all of the superfields
 so as  to elegantly embed 
      the original equations of motion in a bigger  framework. We saw that in some situations our glueing conditions do not satisfy the cocycle condition on 
        triple intersections 
        so that the space is not properly a manifold, but formally these spaces appear to have well-defined complex and symplectic structures.

 A remarkable construction of generalised K\"ahler geometry of symplectic type was given in  \cite{Bischoff:2018kzk} and one of the motivations of this work was to try find a  physicists' explanation of this. This and related work by Gualtieri  and collaborators \cite{Marco-paper} involves realising the generalised K\"ahler geometry as a subspace of a  high dimensional space.
 An important recent development is the construction of   a generic generalised K\"ahler manifold from a space of quadrupled dimension;
 see
    \cite{Marco-paper} and  PhD thesis \cite{jiang}.    
 Specifically, a holomorphic symplectic double Morita bimodule (quadrupling the original dimension) is constructed. This has a real structure such that the fixed point locus is real symplectic (double the original dimension) and  has a Lagrangian submanifold (of the same dimension as the original manifold) which determines the metric.
 Our construction of  generalised K\"ahler geometry also involves quadrupling some or all of the dimensions.
 It will be very interesting to investigate the relation between these two quadrupling constructions.

   One   drawback of our   discussion is that it assumes the regularity of the Poisson foliations appearing in the generalised K\"ahler 
     geometry. We need to impose this requirement in order to use the original superfield formalism involving different kind of fields (chiral, 
      twisted chiral and semi-chiral)  globally. 
      However, in general this is not the case:  generalised K\"ahler 
     geometry can exhibit type change so that
     different regions of the space will have different foliation structures \cite{Gualtieri:2003dx}.
      In Section \ref{s:double} we reformulated the equations of motion  in terms of semi-chiral fields only with a quadrupled target space.
        In this formulation the chirality conditions are emergent, arising from 
          imposing a real Lagrangian condition via the real generating function $\hat{K}$ defined
         on the bigger space. We can imagine a situation in which the changes in $\hat{K}$  lead to 
            changes in the superspace constraints    (corresponding to type
          change). 
          That is, a formulation in such an extended space could provide a global description even for target spaces in which there is type change.
          We find this observation exciting and   plan to study it further elsewhere.

         Our new approach suggests many other future avenues of research. One is to investigate further the  relation between the dual formulations of the sigma model with particular attention to the global properties and the issue of whether the dual theories are fully equivalent classically, and then to investigate the relation between the corresponding quantum theories. 
           It will be interesting to understand the implications of our approach for models with (2,0), (2,1), (4,2) or (4,4) supersymmetry.
           Much remains to be understood about the global geometry of generalised K\" ahler spaces and how to construct them and our formulation provides a new approach to these issues.

\bigskip\bigskip
\noindent{\bf\Large Acknowledgement}:
\bigskip

\noindent We are grateful to Francesco Bonechi and Marco Gualtieri for illuminating discussions. 
We thank the Galileo Galilei Institute
for Theoretical Physics, Florence for providing the stimulating atmosphere where this work was initiated.
 The research of CH is supported by   the STFC Consolidated Grants ST/T000791/1 and ST/X000575/1.
  The research of MZ is  supported by the VR excellence center grant ``Geometry and Physics'' 2022-06593.

\appendix

\section{Appendix: $N=(1,1)$ and $N=(2,2)$ superspace conventions}

Here we sumarise briefly our 
superspace conventions. We follow  the conventions used in \cite{Hitchin:1986ea} and  in
 \cite{Lindstrom:2005zr}. For a detailed introduction to superfield formalism, see \cite{Gates:1983nr}.

We work in 2-dimensional Minkowski  space and use two-component spinors $\psi^\alpha$ which carry a spinor index $\alpha=(+,-)$.
 The spinor indices are raised and lowered using the antisymmetric charge conjugation matrix $C_{\alpha\beta}$ with
 \be C_{+-}=i~,
 \qquad
 C_{\alpha\beta}= - C_{\beta\alpha}= - C^{\alpha\beta}~,
 \ee
 so that
 \be
 \psi ^\alpha = C^{\alpha\beta}\psi_\beta~,
 \qquad
\psi _\alpha = \psi^\beta C_{\beta \alpha}~,
 \qquad
\psi _\pm =\mp i \psi ^\mp~.
 \ee
 The Lorentz-invariant spinor inner product is
 \be
 C_{\alpha\beta}\psi ^\alpha \chi ^ \beta=
 \psi _\alpha \chi ^ \alpha
 =
 i\psi ^+
\chi^--
i\psi ^-
\chi^+
 \ee
 and its complex conjugate is
 \be
 (C_{\alpha\beta}\psi ^\alpha \chi ^ \beta)^*=
- i(\psi ^+)^*
(\chi^-)^*
+
i(\psi ^-)^*
(\chi^+)^*=
C_{\alpha\beta}( \chi ^ \alpha )^* (\psi ^\beta)^*
 \ee
 (where we have assumed that both $\psi$ and $\chi$ are Grassman odd).
 
 The $N=(1,1)$ superspace $\mathbb{R}^{2|2}$ has an even sector which is two-dimensional Minkowski space with coordinates $x^{\alpha\beta}$ which we represent as (symmetric traceless)  bispinors, and we use the notation
 $x^{\+}=x^{++}, x^==x^{--}$.
 The  corresponding even derivative is written as $\partial_{\alpha\beta} = \frac{\partial}{\partial x^{\alpha\beta}}$.
The Grassmann (odd) coordinates $\theta^\alpha$ 
are real  (Majorana) two component spinors.
    On $\mathbb{R}^{2|2}$ we  define two  odd first order differential operators $D_\alpha$ and $Q_\alpha$
    as follows
 \bea
 D_+ = \partial_{+} + i \theta^{+} \partial_{\+}~,~~~~~D_- = \partial_- + i \theta^- \partial_{=}~, \\
  Q_+ =  i \partial_{+} +  \theta^{+} \partial_{\+}~,~~~~~Q_- =  i\partial_- +  \theta^- \partial_{=}~.
 \eea
 They 
   satisfy the  algebra \bea
 D_+^2 = i \partial_\+ ~,~~~~D_-^2 = i \partial_=~,~~~~\{ D_+, D_- \}=0~,\\
 Q_+^2 = i \partial_\+ ~,~~~~Q_-^2 = i \partial_=~,~~~~\{ Q_+, Q_- \}=0~,
\eea
with
\be
 \{ Q_\alpha, D_\beta \}=0~.
\ee
In the supersymmetry literature, the $D_\alpha$ are referred to as   spinorial covariant derivatives and the $Q_\alpha$ are referred to as the  supersymmetry generators. 
 On the space of maps
 \be
   \{ \mathbb{R}^{2|2}~\longrightarrow ~M\}
 \ee
 $Q_\alpha$ and $D_\alpha$ can be viewed as odd vector fields satisfying the algebra presented above. 

 Next we discuss the  $N=(2,2)$ superspace  $\mathbb{R}^{2|4}$. 
 The bosonic coordinates are $x^{\alpha\beta}$ as before, but now the
 grassmann variables are \emph{complex} spinors $\theta^\alpha$ with complex conjugates $\bar{\theta}^\alpha$ and odd derivatives $(\partial_\alpha, \bar{\partial}_\alpha)$.
  We  define two sets of first order odd differential operators $(\mathbb{D}_\alpha, \bar{\mathbb{D}}_\alpha)$ and $(\mathbb{Q}_\alpha, 
  \bar{\mathbb{Q}}_\alpha)$
  by
  \bea
 \mathbb{D}_+ = \partial_{+} + \frac{i}{2} \bar{\theta}^{+} \partial_{\+}~,~~~~~\bar{\mathbb{D}}_+ = \bar{\partial}_+ + \frac{i}{2} \theta^+
  \partial_{\+}~, \\
 \mathbb{Q}_+ = i \partial_{+} + \frac{1}{2} \bar{\theta}^{+} \partial_{\+}~,~~~~~\bar{\mathbb{Q}}_+ = i \bar{\partial}_+ + \frac{1}{2}  \theta^+
  \partial_{\+}~, 
 \eea
 for the $+$ sector, with analogous formulae for the  $-$ sector. 
The algebra is given by the following relations
\bea\label{N22-algebra}
 \{ \mathbb{D}_+, \bar{\mathbb{D}}_+\} = i \partial_{\+}~,~~~~ \{ \mathbb{D}_-, \bar{\mathbb{D}}_-\} = i \partial_{=}~,~~~~
\{ \mathbb{D}_\alpha, \mathbb{D}_\beta \}=0~,~~~~\{ \bar{\mathbb{D}}_\alpha, \bar{\mathbb{D}}_\beta \}=0~,\\
 \{ \mathbb{Q}_+, \bar{\mathbb{Q}}_+\} = i \partial_{\+}~,~~~~ \{ \mathbb{Q}_-, \bar{\mathbb{Q}}_-\} = i \partial_{=}~,~~~~
  \{ \mathbb{Q}_\alpha, \mathbb{Q}_\beta \}=0~,~~~~\{ \bar{\mathbb{Q}}_\alpha, \bar{\mathbb{Q}}_\beta \}=0~,
\eea
 where all $\mathbb{Q}/\bar{\mathbb{Q}}$-operators anti commute with all $\mathbb{D}/\bar{\mathbb{D}}$-operators. 
 Again, in the supersymmetry literature,  $\mathbb{D}_\alpha, \bar{\mathbb{D}}_\alpha$ are referred to as   spinorial covariant derivatives and $\mathbb{Q}_\alpha, \bar{\mathbb{Q}}_\alpha$ are referred to as    supersymmetry generators. 
 On the space of maps  
 \be
\{ \mathbb{R}^{2|4}~\longrightarrow ~M\}
 \ee
$\mathbb{D}_\alpha, \bar{\mathbb{D}}_\alpha$  and $\mathbb{Q}_\alpha, \bar{\mathbb{Q}}_\alpha$ can be interpreted as 
 odd vector fields with the algebra given above. 
 
 Here we have a new feature, we can consider maps which are annihilated by 
  specific nilpotent combinations of $\mathbb{D}$'s and $\bar{\mathbb{D}}$'s so that  the supersymmetry algebra generated by $\mathbb{Q}$, $\bar{\mathbb{Q}}$
   acts on the subspace that is constrained in this way. That is, the superfields constrained in this way furnish a representation of the supersymmetry algebra generated by 
   $\mathbb{Q},\bar{\mathbb{Q}}$ and $ \frac{\partial}{\partial x^{\alpha\beta}}$.
    Below we summarise the  constraints,
    that we use throughout the paper
 \begin{center}
\begin{tabular}{|c|c|}
\hline
    $\bar{\mathbb{D}}_\pm \phi =0$ &  chiral superfield   \\
\hline $\bar{\mathbb{D}}_+  \bar{\mathbb{D}}_- \Sigma =0$ &  complex linear superfield    \\
\hline  $\bar{\mathbb{D}}_+ \chi=0~,~~~\mathbb{D}_- \chi=0$  &   twisted chiral fields \\
\hline   $\bar{\mathbb{D}}_+  \mathbb{D}_- \Lambda =0$ &  twisted complex linear superfield    \\
 \hline
$\bar{\mathbb{D}}_+ \bar{X}_L =0$  (or $\bar{\mathbb{D}}_+ \bar{Y}_L =0$) &  left semichiral superfield   \\
  \hline
$\bar{\mathbb{D}}_- X_R =0$  (or $\bar{\mathbb{D}}_- Y_R =0$) &  right semichiral superfield   \\
 \hline
\end{tabular}
\end{center}
 Each of these constraints has a general solution in terms of unconstrained complex superfields. These are as follows:
  \bea
 && \phi = \bar{\mathbb{D}}_+ \bar{\mathbb{D}}_- \Phi \nn \\
  &&\Sigma =  \bar{\mathbb{D}}_+ \Phi_1 + \bar{\mathbb{D}}_- \Phi_2  \nn \\
   && \chi= \bar{\mathbb{D}}_+ \mathbb{D}_- \Phi  \label{reps-super} \\
   &&\Lambda =  \bar{\mathbb{D}}_+ \Phi_1 + \mathbb{D}_- \Phi_2 \nn \\
   && X_L = \bar{\mathbb{D}}_+ \Phi \nn  \\
     && X_R = \bar{\mathbb{D}}_- \Phi  \nn
  \eea
  Here the $\Phi$'s are unrestricted complex superfields.

 The constrains on the superfields allow the following 
  list of field redefinitions which are compatible with the constraints 
 \bea
&& \phi~\longrightarrow~f_1(\phi)~, \nn \\
&& \Sigma~\longrightarrow~f_2(\phi)\Sigma  + f_3 (\phi, \chi, X_L) + f_4(\phi, \bar{\chi}, X_R) ~,\nn \\
 &&\chi~\longrightarrow~f_5(\chi)~,\label{superf-transf} \\
 && \Lambda ~\longrightarrow~ f_6(\chi) \Lambda + f_7 (\phi, \chi, X_L) + f_8 (\bar{\phi}, \chi, \bar{X}_R)  ~,\nn \\
 && X_L ~\longrightarrow~ f_9(\phi, \chi, X_L)~, \nn \\
&& X_R~\longrightarrow~f_{10}(\phi, \bar{\chi}, X_R)~,\nn
 \eea
  (together with their complex conjugates) and  here $f_i$ are arbitrary functions.  In the context of non-linear sigma models
   the lowerest components of these superfields are interpreted as coordinates on the target space $M$ and thus we get the restrictions on 
    the diffeomorphisms that are compatible with the constraints, and these in turn are related to  geometric structures on $M$. 
    
 \section{Appendix: Generalised K\"ahler geometry, gerbes and glueings}   
 
 In this appendix we review the relevant facts about  generalised K\"ahler geometry and gerbes. We follow closely  \cite{Hull:2008vw} and point out some properties which were not discussed in \cite{Hull:2008vw} but which are important  for this paper. 
 We consider a smooth manifold $M$ with an open cover $\{ U_\alpha \}$ such that all open sets and finite  intersections thereof are contractible.
 
 A set of 
 transition functions $f_{\alpha\beta}:U_\alpha \cap U_\beta \to \mathbb{R}$ with  ($f_{\alpha\beta}=-f_{\beta\alpha}$) from the intersection of  any two patches $U_\alpha \cap U_\beta$ to 
 $\mathbb{R}$
 define a real line bundle if they satisfy the cocycle condition on triple overlaps $U_\alpha \cap U_\beta \cap U_\gamma$
 that
 \be
  f_{\alpha\beta}+f_{\beta\gamma}+f_{\gamma\alpha}=0
   \ee
   while 
   if they satisfy this modulo $2 \pi$ then
   the maps
   $g_{\alpha\beta}:U_\alpha \cap U_\beta \to U(1)$ 
   with $
       g_{\alpha\beta}=e^{if_{\alpha\beta}}$  
      satisfy the cocycle condition
\be  g_{\alpha\beta}g_{\beta\gamma}g_{\gamma\alpha}=1
   \ee
    on $U_\alpha \cap U_\beta \cap U_\gamma$ and so
    define a complex line bundle. A gerbe is a set of maps on triple intersections 
    $h_{\alpha\beta\gamma}:U_\alpha \cap U_\beta \cap U_\gamma \to \mathbb{R}$ satisfying
    \be
        h_{\beta\gamma\delta} +h_{\delta\gamma\alpha}+h_{\alpha\beta\delta}+h_{\beta\alpha\gamma}
        =0
        \ee
        modulo $2\pi$ so that
         $k_{\alpha\beta\gamma}=e^{ih_{\alpha\beta\gamma}}$ are maps
          $k_{\alpha\beta\gamma}:U_\alpha \cap U_\beta \cap U_\gamma \to U(1)$
          satisfying
          \be
        k_{\beta\gamma\delta} k_{\delta\gamma\alpha}k_{\alpha\beta\delta}k_{\beta\alpha\gamma}
        =1
        \label{gerby}
        \ee
  
  We follow the setup and the notation of \cite{Hull:2008vw}. On the intersection of two patches $U_\alpha \cap U_\beta$
   the generalized real K\"ahler potential has the glueing
   \be
       K_\alpha - K_\beta = F_{\alpha\beta}^+ (\phi, \chi, X_L) + \bar{F}^+(\bar{\phi}, \bar{\chi}, \bar{X}_L) + 
     F_{\alpha\beta}^- (\phi, \bar{\chi}, X_R) + \bar{F}_{\alpha\beta}^-(\bar{\phi}, \chi, \bar{X}_R)~.
 \ee
  The functions $F^\pm$ are defined up to the following ambiguities
  \bea
   && F_{\alpha\beta}^+ (\phi, \chi, X_L) ~\rightarrow~F_{\alpha\beta}^+ (\phi, \chi, X_L) + \rho_{\alpha\beta} (\phi) + \sigma_{\alpha\beta}(\chi)~,\\
   &&  F_{\alpha\beta}^- (\phi, \bar{\chi}, X_R) ~\rightarrow~  F_{\alpha\beta}^- (\phi, \bar{\chi}, X_R) - \rho_{\alpha\beta} (\phi) - \bar{\sigma}_{\alpha\beta}(\bar{\chi})~
  \eea
and they  can be chosen such that $F_{\alpha\beta}^\pm = - F_{\beta\alpha}^\pm$. Moreover they satisfy the following 
   conditions on the triple intersections $U_\alpha \cap U_\beta \cap U_\gamma$
   \bea
 &&  F_{\alpha\beta}^+ (\phi, \chi, X_L) +  F_{\beta\gamma}^+ (\phi, \chi, X_L) +  F_{\gamma\alpha}^+ (\phi, \chi, X_L) =i c_{\alpha\beta\gamma}(\phi) - i b_{\alpha\beta\gamma}(\chi)~, \\
 &&   F_{\alpha\beta}^- (\phi, \bar{\chi}, X_R) +  F_{\beta\gamma}^- (\phi, \bar{\chi}, X_R) +  F_{\gamma\alpha}^- (\phi, \bar{\chi}, X_R) 
    = - i c_{\alpha\beta\gamma}(\phi) - i  \bar{b}_{\alpha\beta\gamma}(\bar{\chi}) ~
   \eea
   for some functions $c_{\alpha\beta\gamma}(\phi) ,   {b}_{\alpha\beta\gamma}( {\chi}) $.
    On quadruple intersections $U_\alpha\cap U_\beta \cap U_\gamma \cap U_\delta$  
    \bea
   &&  c_{\beta\gamma\delta} (\phi) + c_{\delta\gamma\alpha}(\phi) + c_{\alpha\beta\delta}(\phi) + c_{\beta\alpha\gamma}(\phi) = \frac{i}{4} d_{\alpha\beta\gamma\delta}~,\\
    &&  b_{\beta\gamma\delta} (\chi) + b_{\delta\gamma\alpha}(\chi) + b_{\alpha\beta\delta}(\chi) + b_{\beta\alpha\gamma}(\chi) = \frac{i}{4} d_{\alpha\beta\gamma\delta}~,\
    \eea
     where $d_{\alpha\beta\gamma\delta}$ is constant. If the 3-form $H/2\pi$ is quantised, i.e.\  if it represents an integral cohomology class, then $d_{\alpha\beta\gamma\delta} \in 2\pi \mathbb{Z}$   
     so that $e^{4c_{\alpha\beta\gamma}}$ and $e^{4b_{\alpha\beta\gamma}}$ 
     each satisfy
    (\ref{gerby}) and so each defines a gerbe, and since these are holomorphic functions, they are holomorphic gerbes.
     In our discussion here of the classical theory, the quantisation condition on $H$ does not play a  role.
      
    Using the above properties we can construct different objects which satisfy the cocycle condition on the triple intersection. For example we have the following 
     relation on   $U_\alpha \cap U_\beta \cap U_\gamma$
     \bea
     \frac{\partial}{\partial \phi^a} \Big ( F_{\alpha\beta}^+ (\phi, \chi, X_L)  +   F_{\alpha\beta}^- (\phi, \bar{\chi}, X_R)  \Big ) +
     \frac{\partial}{\partial \phi^a} \Big ( F_{\beta\gamma}^+ (\phi, \chi, X_L)  +   F_{\beta\gamma}^- (\phi, \bar{\chi}, X_R)  \Big ) \nn \\
     +\frac{\partial}{\partial \phi^a} \Big ( F_{\gamma\alpha}^+ (\phi, \chi, X_L)  +   F_{\gamma\alpha}^- (\phi, \bar{\chi}, X_R)  \Big )=0~.
     \eea
     Analogously, the following objects satisfy the cocycle conditions on triple intersections $U_\alpha\cap U_\beta \cap U_\gamma$ 
     \bea
   &&  \frac{\partial}{\partial \chi^{a'}} \Big ( F_{\alpha\beta}^+ (\phi, \chi, X_L)  + \bar{F}_{\alpha\beta}^-(\bar{\phi}, \chi, \bar{X}_R) \Big )~,   \\
       &&  \frac{\partial}{\partial X^n_L} F_{\alpha\beta}^+ (\phi, \chi, X_L) ~,  \\
          &&   \frac{\partial}{\partial X^{n'}_R} F_{\alpha\beta}^- (\phi, \bar{\chi}, X_R) ~.
     \eea
    Thus we can use them to construct different affine bundles. 
    
    Next we turn to the geometrical setting of the construction of section 5.
    Let $\partial _\pm
    ,\bar \partial _\pm$ be the Dolbeaut exterior derivatives for the complex structure $I_\pm$.
    Suppose in each $U_\alpha$ there is a (1,0) form
    $X_\alpha$ with
    \be
    X_\alpha= X_{\alpha a }dz^a+X_{\alpha a'}dr^{a'} +X_{\alpha n}d\zeta^n
    \ee
    with glueing relations
    \be
    X_\beta=X_\alpha+ \partial _+ F^+_{\alpha \beta}
    \ee
    These then satisfy (B.8),(B.10),(B.11) and so see the gerbe structure.
    Defining $b_ \alpha=\bar \partial _+ X_\alpha$,
    we have
    \be
    b_\beta=b_\alpha
    \ee
    so $b$ is a globally defined (1,1) form and
    \be 
    h\equiv 
    \bar \partial _+ b_\alpha
    \ee
    is zero, so that $b_\alpha$ is a holomorphic gerbe connection which is flat ($h=0$), with prepotential $X_\alpha$.
 Now    the components
 $(X_{\alpha a },X_{\alpha a'},X_{\alpha n}
 )$ have exactly the same glueing relations  as $p_L,r_L,\pi_L$ from
 (\ref{gen-shift-9}),(\ref{gen-shift-10}), i.e. $p_L,r_L,\pi_L$ can be viewed as the components of the prepotential for a flat holomorphic gerbe connection.

\bibliographystyle{utphys}
\bibliography{references}{}

\end{document}